\RequirePackage{fix-cm}
\documentclass[twocolumn,epjc3]{svjour3}  
\smartqed  
\usepackage[dvipdfmx]{graphicx}
\RequirePackage[T1]{fontenc}
\RequirePackage{mathptmx} 

\usepackage{bm}
\usepackage{float}
\usepackage{here}
\usepackage{enumerate}
\usepackage{makeidx}
\usepackage{extarrows}
\usepackage{fancybox}
\usepackage{fancyhdr}
\usepackage{indentfirst}
\usepackage{cite}
\usepackage{color}

\journalname{Eur. Phys. J. C}
\begin{document}

\title{A new relativistic hydrodynamics code for high-energy heavy-ion collisions
}


\author{Kazuhisa Okamoto \thanksref{addr1,e1}
        \and  Yukinao Akamatsu \thanksref{addr2, addr3, addr4} 
        \and Chiho Nonaka \thanksref{addr1, addr2, addr5}
}

\thankstext{e1}{e-mail: okamoto@hken.phys.nagoya-u.ac.jp}


\institute{Department of Physics, Nagoya University, Nagoya 464-8602, Japan \label{addr1}
           \and Kobayashi-Maskawa Institute for the Origin of Particles and the Universe (KMI), Nagoya University, Nagoya 464-8602, Japan \label{addr2}
           \and  Department of Physics, Osaka University, Toyonaka 560-0043, Japan  \label{addr3}
	   \and Department of Physics and Astronomy, Stony Brook University, Stony Brook, NY 11794, USA \label{addr4}
           \and Department of Physics, Duke University, Durham, NC 27708, USA \label{addr5}     
}
\date{Received: date / Accepted: date}
\maketitle
\begin{abstract}
We construct a new Godunov type relativistic hydrodynamics code in Milne coordinates,
using a Riemann solver based on the two-shock approximation which is stable under the existence of large shock waves.
We check the correctness of the numerical algorithm by comparing numerical calculations and analytical solutions in various problems, such as shock tubes, expansion of matter into the vacuum, the Landau-Khalatnikov solution, and propagation of fluctuations around Bjorken flow and Gubser flow. 
We investigate the energy and momentum conservation property of our code in a test problem of longitudinal hydrodynamic expansion with an initial condition for high-energy heavy-ion collisions. 
We also discuss numerical viscosity in the test problems of expansion of matter into the vacuum and conservation properties.  
Furthermore, we discuss how the numerical stability is affected by the source terms of relativistic numerical hydrodynamics in Milne coordinates.

\end{abstract}

\section{Introduction}
Relativistic hydrodynamics has been widely used for the description of macroscopic dynamics in various fields ranging from nuclear physics to astrophysics.
The high-energy heavy-ion collision experiment is one of the active areas of relativistic hydrodynamics applications.

In 2005 at the relativistic heavy-ion collider (RHIC), the production of strongly interacting quark-gluon plasma (QGP) was achieved, which was supported not only by the experimental data but also theoretical analyses \cite{QGP_RHIC}. 
Studies based on relativistic hydrodynamics have shown remarkable success in understanding various observables such as particle distributions, collective flows, particle correlations, and so on \cite{deSouza:2015ena,Gale:2013da,Hirano:2012kj,NA}. 
The strong elliptic flow at RHIC is a highlight of the success of hydrodynamic models and is one piece of evidence that the QGP is not a weakly interacting gas but a strongly interacting matter. 
Since then the hydrodynamic model has been one of the promising phenomenological models for the description of dynamics of hot and dense matter produced in the heavy-ion collisions. 

The construction of relativistic viscous hydrodynamic model has been of practical importance and has formed a basis for the analyses of the heavy-ion collisions \cite{Song:2007ux,Baier:2006gy,Dusling:2007gi,Denicol:2009am,Schenke:2010rr,Bozek:2011ua,Pang:2014ipa}. 
In the last decade, the hydrodynamic model itself has also been developed through the analyses of experimental data of heavy-ion collisions at RHIC and the large hadron collider (LHC).  
By comparing the hydrodynamic model calculations and the experimental observables such as particle distributions and collective flows, detailed bulk properties of QGP such as the QCD equation of state and its transport coefficients have been investigated.  
Also, physical QCD equation of state is now available by lattice QCD simulations at vanishing chemical potential \cite{Lattice1, Lattice2} and is applied to the hydrodynamic model.
This enables us to bridge the first principle lattice QCD simulations and the experimental data in the heavy-ion collisions.

In spite of the success of hydrodynamic models in high-energy heavy-ion collisions, there are still several issues under discussion. 
Currently the hydrodynamic models often adopt Israel-Stewart theory \cite{IsSt2} and a second-order viscous hydrodynamics from AdS/CFT correspondence \cite{Baier:2007ix} as their basic equations. 
However, we have not reached a conclusion on which relativistic viscous hydrodynamic equation is suitable for the description of relativistic heavy-ion collisions.
This is because the extension from a relativistic ideal hydrodynamic equation to a viscous hydrodynamic equation is not straightforward and several possible candidates exist.
Also viscous second-order anisotropic hydrodynamics is proposed which reproduces the exact solution of the Boltzmann equation in the relaxation-time approximation \cite{Bazow:2013ifa}. 
Furthermore it remains an enormous challenge to understand why hydrodynamics can be applied to the dynamics shortly after a heavy-ion collision takes place.
Conclusive understanding of the mechanism of thermalization and hydrodynamization on such a short time scale is still missing.

Here we emphasize that a numerical algorithm for solving the relativistic hydrodynamic equation is one of the important ingredients in developing the hydrodynamic models.
Recent high statistical experimental data at RHIC and the LHC imposed a more rigorous numerical treatment on the hydrodynamical models.
For example, at RHIC and the LHC, higher harmonic anisotropic flow, which is expressed by the higher Fourier coefficient of particle yields as a function of azimuthal angle, is reported \cite{vn-PHENIX, vn-STAR, vn-ALICE, vn-ATLAS, vn-CMS}. 
The origin of the higher harmonics is considered to be event-by-event initial fluctuations in the particle distributions. 
When comparing with those high statistical data, reducing the numerical dissipation of the numerical algorithm for relativistic hydrodynamic equations should allow us an access to more precise value of transport coefficients of the QGP.
Usually each algorithm has advantages or disadvantages in terms of coding, computational time, numerical precision, and stability. 
Up to now, unfortunately, only little attention has been paid to the numerical aspects in the hydrodynamic models for high-energy heavy-ion collisions.

Recently we developed a state-of-the-art numerical algorithm for solving the relativistic hydrodynamic equation with the QGP equation of state \cite{Akamatsu2014}. 
In the algorithm, we use a Riemann solver based on the two-shock approximation \cite{Colella1982,Balsara1994,Dai1997,Mignone2005} which is stable under the existence of large shock waves \cite{Woodward1984}. 
The new numerical scheme is stable even with a small numerical viscosity and can reduce the numerical uncertainly when extracting the physical viscosity of the QGP from the experimental data. 
However, this algorithm in Ref. \cite{Akamatsu2014} is developed in Cartesian coordinates. 
Meanwhile, at the high-energy heavy-ion collisions such as RHIC and the LHC, the expansion in longitudinal direction is rapid compared with that in transverse direction.
For the description of a space-time evolution of high-energy heavy-ion collisions, Milne coordinates are more suitable than Cartesian coordinates.
Therefore we extend our algorithm of relativistic ideal hydrodynamics in Cartesian coordinates to that in Milne coordinates so that we can efficiently apply it to the analyses of high-energy heavy-ion collisions.
The algorithm that we shall present here plays an important role 
in solving the relativistic viscous hydrodynamic equation numerically \cite{Inutsuka2011, Akamatsu2014}.
For the viscous hydrodynamics, we split the hydrodynamic equations into an ideal part and a viscous part. The ideal part can be solved by the Riemann solver for ideal hydrodynamics.

The present article is organized as follows. 
We begin in Sect.\:2 by showing the basic equations for the hydrodynamic models in Milne coordinates. 
In Sect.\:3 we explain the numerical algorithm; Riemann problem in Milne coordinates and our numerical scheme. 
Section 4 is devoted to several numerical tests, such as relativistic shock tubes and a comparison with analytic solutions which describe the dynamics of realistic high-energy heavy-ion collisions. 
In addition, we discuss the propagation of longitudinal fluctuations.
In Sect.\:5 we investigate the conservation property of our code. 
We end in Sect.\:6 with our conclusions. 

\section{Relativistic hydrodynamics}
Relativistic hydrodynamics is based on the conservation equations of net charge, energy, and momentum,  
\begin{align}
& J^\mu_{\; ; \mu}=0 \label{eq:cons1},\\
&T^{\mu\nu}_{\;\;\; ;\mu}=0 \label{eq:cons2},
\end{align}
where $J^{\mu}$ is the baryon number current and $T^{\mu\nu}$ is the energy-momentum tensor.
For the ideal fluid, the energy-momentum tensor and the baryon number current are given by
\begin{align}
J^\mu&=n u^\mu, \label{eq:current} \\
T^{\mu\nu}&=(e+p)u^\mu u^\nu - p g^{\mu\nu},\label{eq:tensor}
\end{align}
where $n$ is the baryon number density, $e$ is the energy density, $p$ is the pressure, $u^\mu$ is the normalized four-velocity of the fluid, $u^\mu u_\mu=1$ and $g^{\mu\nu}$ is the metric tensor.

\par In high-energy heavy-ion collisions at RHIC and the LHC, approximate invariance under the longitudinal Lorentz boost is observed in particle rapidity distributions around mid-rapidity \cite{BRAHMS2001, BRAHMS2002, PHOBOS2003, PHOBOS2011, CMS2011, ATLAS2012, ALICE2013}.
In such situations, Milne coordinates are suitable for the description of  space-time evolution of the hot and dense matter after the collisions.
Milne coordinates $\eta$ and $\tau$ are described by the rapidity $\eta ={\rm tanh}^{-1} (z/t)$ and the proper time $\tau = \sqrt{t^2 -z^2}$ with Cartesian coordinates. 
The coordinate transformation of the four-velocity between Milne coordinates and Cartesian coordinates is 
given by 
\begin{align} u^\tau &={\rm cosh}\eta\: u^t -{\rm sinh}\eta\: u^z \label{eq:trans-utau}, \\
u^\eta &=-\frac{{\rm sinh}\eta}{\tau}u^t+\frac{{\rm cosh}\eta}{\tau}u^z \label{eq:trans-ueta}, 
\end{align} 
where the transverse components of the four-velocity, $u^x$ and $u^y$ are the same in both coordinates. 
The four-velocity in Milne coordinates is written by three-dimensional velocity as
\begin{equation} u^\alpha=(u^\tau, u^x, u^y, u^\eta)=W(1, w^x, w^y, w^\eta), 
\end{equation}
where $w^i=u^i/u^\tau(i=x,y,\eta)$ is the three-dimensional velocity in Milne coordinates and  $W$ represents the Lorentz factor,
\begin{equation}
W=u^\tau=\left( 1- (w^x)^2 - (w^y)^2 - \tau^2 (w^\eta)^2\right)^{-1/2}.
\end{equation}
The coordinate transformation of the three-dimensional velocity vector between Milne coordinates and Cartesian coordinates is given by 
\begin{align}
w^\eta &= \frac{1}{\tau}\frac{-{\rm sinh}\eta +v^z{\rm cosh}\eta }{{\rm cosh}\eta - v^z{\rm sinh}\eta }, \label{eq:trans-veta}  \\  
w^i &=\frac{v^i}{{\rm cosh}\eta -v^z {\rm sinh}\eta},  \hspace{1cm}(i=x, y). \label{eq:trans-vx}
\end{align}
where $v^i = u^i/u^t (i=x,y,z)$ is the three-dimensional velocity in Cartesian coordinates.
In contrast to $u^x$ and $u^y$, $x$ and $y$ components of the three-dimensional velocity in Milne coordinates are different from those in Cartesian coordinates. 
The metric tensor is given by $g^{\alpha\beta}={\rm diag}(1, -1,-1,-1/\tau^2)$ and the nonzero components of the Christoffel symbols are 
\begin{equation}
\Gamma^{\eta}_{\eta\tau}=\Gamma^\eta_{\tau\eta}=\frac{1}{\tau},\qquad \Gamma^\tau_{\eta\eta}=\tau.
\label{eq:metric_m}
\end{equation}

\par In Milne coordinates, the charge conservation equation Eq.\:\eqref{eq:cons1} and the equation of energy and momentum conservation 
Eq.\:\eqref{eq:cons2} are written by 
\begin{align}
&\partial_\tau J^\tau  + \partial_i J^i+\partial_\eta J^\eta= -J^\tau/\tau \label{eq:source1}, \\
& \partial_\tau T^{\tau j} + \partial_i T^{ij} +\partial_\eta T^{\eta j} = - T^{\tau j}/\tau , \label{eq:source2}\\
 &\partial_\tau T^{\tau\eta} +\partial_i  T^{i\eta}+\partial_\eta T^{\eta\eta}  = -3 T^{\tau\eta}/\tau ,\label{eq:source3}\\
 &\partial_\tau T^{\tau\tau}+\partial_i T^{i\tau}+\partial_\eta T^{\eta\tau}= -T^{\tau\tau}/\tau-\tau T^{\eta\eta} \label{eq:source4}.
\end{align}
There are geometric source terms in the right-hand side of Eqs.\:\eqref{eq:source1}-\eqref{eq:source4}, which contain the effect from the coordinate expansion with $\tau$. 
One can rewrite Eqs.\:\eqref{eq:source1}-\eqref{eq:source4} 
\begin{align}
&\partial_\tau (\tau J^\tau ) + \partial_i(\tau J^i)+\partial_\eta(\tau J^\eta)=0  \label{eq:1},\\
& \partial_\tau (\tau T^{\tau j}) + \partial_i (\tau T^{ij}) +\partial_\eta (\tau T^{\eta j}) =0 \label{eq:2},\\
&\partial_\tau (\tau T^{\tau\eta} )+\partial_i (\tau T^{i\eta}) +\partial_\eta (\tau T^{\eta\eta} ) = -2 T^{\tau\eta} \label{eq:3},\\
&\partial_\tau(\tau T^{\tau\tau})+\partial_i (\tau T^{i\tau})+\partial_\eta(\tau T^{\eta\tau})= -\tau^2 T^{\eta\eta} \label{eq:4}.
\end{align}
Here the effect from the coordinate expansion with $\tau$ is absorbed into the Jacobian $\tau$ in the derivative terms.
There are the source terms in the right-hand side of Eqs.\:\eqref{eq:3} and \eqref{eq:4}, which indicates that $T^{\tau\tau}$ and $T^{\tau\eta}$ are not the conserved quantities.
Instead of them, the conserved quantities are $T^{\tau t}$ and $T^{\tau z}$.
$T^{\tau t}$ and $T^{\tau z}$ are related to $T^{\tau\tau}$ and $T^{\tau\eta}$ through the coordinate transformation between Milne and Cartesian coordinates,
\begin{align}
T^{\tau t}&= {\rm cosh}\eta \:T^{\tau\tau}+\tau {\rm sinh}\eta \:T^{\tau\eta},\\
T^{\tau z}&= {\rm sinh}\eta\: T^{\tau\tau}+\tau{\rm cosh}\eta \: T^{\tau\eta}.
\end{align}
Using the conserved quantities, $T^{\tau t}$ and $T^{\tau z}$, one can express the hydrodynamic equations in the conservative forms \cite{Jeon2015}
\begin{align}
&\partial_\tau (\tau J^\tau ) + \partial_i(\tau J^i)+\partial_\eta(\tau J^\eta)=0, \label{eq:5}\\
&\partial_\tau (\tau T^{\tau j}) + \partial_i (\tau T^{ij}) +\partial_\eta (\tau T^{\eta j}) =0,\label{eq:6}\\
&\partial_\tau (\tau T^{\tau z} )+\partial_i (\tau T^{iz}) +\partial_\eta (\tau T^{\eta z} ) =0, \label{eq:7}\\
&\partial_\tau(\tau T^{\tau t})+\partial_i (\tau T^{i t})+\partial_\eta(\tau T^{\eta t})=0.\label{eq:8}
\end{align}
Here, Eq.\:\eqref{eq:7} corresponds to the conservation of the $z$ component of momentum.
$T^{\tau z}$ represents the density of the $z$ component of momentum at specific proper time. $T^{\eta z}$ represents the flux of the $z$ component of momentum passing through the surface perpendicular to the $\eta$ coordinate. 
Equation \eqref{eq:8} corresponds to energy conservation.
We construct a new algorithm for the relativistic hydrodynamic equations using Eqs.\:\eqref{eq:5}-\eqref{eq:8}, which do not have source terms. 
In numerical tests, Sect.\:\ref{sec:numerical_test}, we shall discuss the effects of the existence of source terms from the point of view of stability and numerical viscosity.

\section{Numerical simulations in Milne coordinates} \label{sec:code}

\subsection{Riemann problem in Milne coordinates} \label{sec:riemann problem}
The Riemann problem is an initial-value problem for the hydrodynamic equation. 
The initial condition is given by two arbitrary constant hydrodynamic  states $\bm V_L$ and  $\bm V_R$ separated by a discontinuity, 
\begin{equation} \bm V(t=t_0, x,y,z)= \left\{
\begin{array}{c} \bm V_L \quad (z<z_i) ,\\
 \bm V_R \quad (z>z_i) ,
\end{array} \right.\label{eq:riemann}
\end{equation}
where $t_0$ and $z_i$ stand for an initial time and a location of the discontinuity, respectively. 
The hydrodynamic state $\bm V= (n,v^x,v^y,v^z, p)$ contains information on fluid variables, namely baryon number density, fluid velocities, and pressure.
The initial discontinuity at $z=z_i$ decays into three nonlinear waves \cite{Marti1994,Pons2000}. 
Two of them are shock waves and/or rarefaction waves. The other is a contact discontinuity moving with hydrodynamic flow. 
These waves evolve between the constant hydrodynamic states $\bm V_L$ and $\bm V_R$ with a constant velocity.
The hydrodynamic states $\bm V_L$ and $\bm V_R$ do not change until characteristic information from the discontinuity arrives. 
Therefore, the hydrodynamic state outside the light cone of the discontinuity $(t_0, z_i)$ remains $\bm V_L$ or $\bm V_R$.

We can also define a Riemann problem in Milne coordinates.
The initial condition of the Riemann problem in Milne coordinates is set to 
\begin{equation}\bm V(\tau=\tau_0, x,y,\eta)= \left\{
\begin{array}{c} \bm V_L \quad (\eta<\eta_i) ,\\
 \bm V_R \quad (\eta>\eta_i) ,
\end{array} \right. \label{eq:initial}
\end{equation}
where $\tau_0$ and $\eta_i$ are the initial proper time and the location of the discontinuity and they represent the same point  as $(t_0, z_i)$ in Eq.\:\eqref{eq:riemann} in Cartesian coordinates. 
Note that the components of $\bm V$ are not $(n,w^x,w^y,w^\eta, p)$ but the same as those in Eq.\:\eqref{eq:riemann}, $(n,v^x,v^y,v^z, p)$. 
The velocity fields in Cartesian coordinates $u^i/u^t=(v^x, v^y, v^z)$ are constant in the rapidity direction.
However, the velocity fields in Milne coordinates $u^i/u^\tau=(w^x, w^y,w^\eta)$ depend on rapidity. 
Now we show that the analytical solution for the Riemann problem in Milne coordinates is obtained from that in Cartesian coordinates by proper coordinate transformations, in which the key issue is to represent the hydrodynamic states as variables independent of $\eta$. 

\begin{figure}[t]
\centering
\includegraphics[width=7cm]{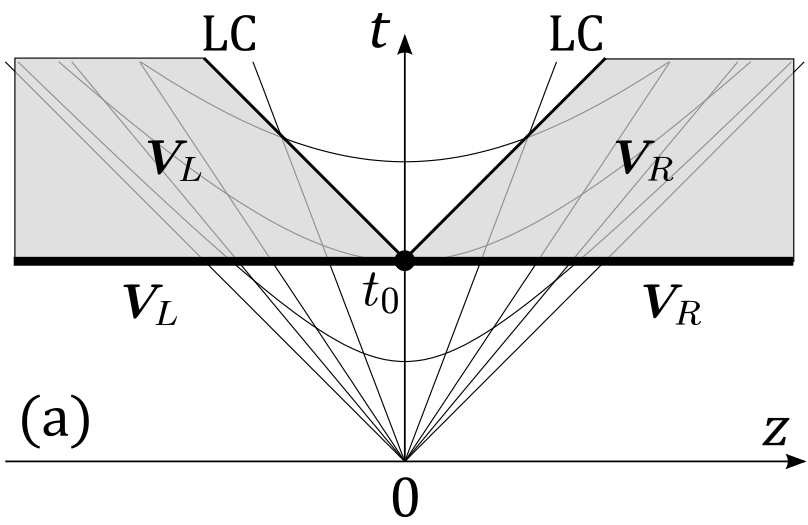}
\centering
\includegraphics[width=7cm]{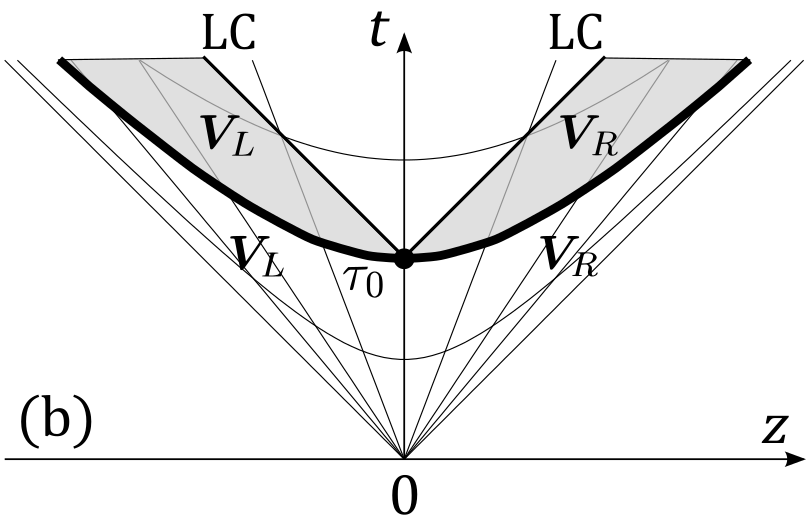}
\caption{
The Riemann problem in Cartesian coordinates (a) and that in Milne coordinates (b). 
The solid circle stands for the initial discontinuity at $z_i=0$ (a) and $\eta_i=0$ (b). 
The thick solid lines indicate the hypersurfaces on which the initial conditions of the Riemann problems are defined. LC stands for the light cone that originates from the discontinuity. \label{fig:initial}}
\end{figure}

Now we compare the two initial-value problems Eqs.\:\eqref{eq:riemann} and \eqref{eq:initial}. 
Without loss of generality, we can assume that the initial discontinuity represented by the solid circle is located at $z_i=\eta_i=0$ as in Figs.\ref{fig:initial}\:(a) and (b). 
In Fig.\:\ref{fig:initial} (a), the thick solid line stands for the time at which we define the initial condition of the Riemann problem in Cartesian coordinates, Eq.\:\eqref{eq:riemann}.
In Milne coordinates, the initial condition of the Riemann problem is set on a hyperbola $\tau=\tau_0$ as in Fig.\:\ref{fig:initial}\:(b).
By comparison between Figs.\ref{fig:initial}\:(a) and (b), the hyperbola $\tau=\tau_0$ in Milne coordinates is located inside the constant hydrodynamic state $\bm V_L$ or $\bm V_R$ in Cartesian coordinates.
This suggests that the initial condition of Eq.\:\eqref{eq:initial} is satisfied by the solution of the Riemann problem in Cartesian coordinates Eq.\:\eqref{eq:riemann}.
In brief, the Riemann problems in both coordinates are identical and thus so are their solutions.
A detailed explanation as regards this proof is given in \ref{app-proof}.

\subsection{Numerical scheme}\label{sec:scheme}

\subsubsection{$\tau-\eta$ system} \label{sec:1d}
\begin{figure}[t]
  \centering
  \includegraphics[width=6.5cm]{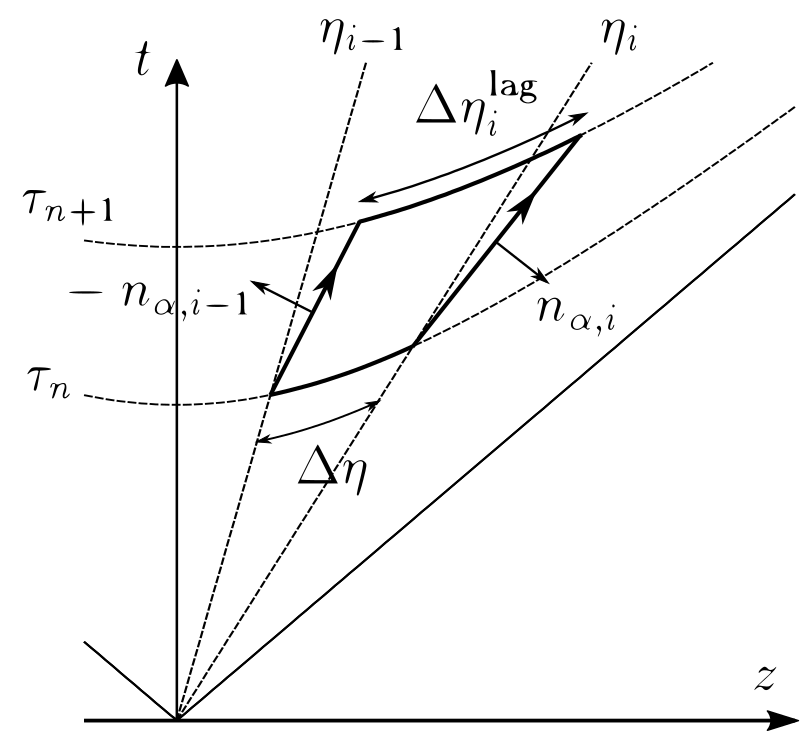}
  \caption{The Lagrange step in Milne coordinates. The solid closed path represents the 
  domain of integration in Eq.\:\eqref{eq:int}. The cell boundary at $(\tau_n, \eta_i)$ moves with fluid.
  $\Delta\eta^{\rm lag}_i\equiv \Delta\eta +(\delta\eta_i(\Delta\tau) -\delta\eta_{i-1}(\Delta\tau))$ is the Lagrange grid-cell size at the proper time $\tau^{n+1}$. 
  See text for detailed explanation. 
  } 
  \label{fig:taueta-lat}
\end{figure}

Assuming that the hydrodynamic variables in the $x$ and $y$ directions are constant, from Eqs.\:\eqref{eq:6}-\eqref{eq:8} we obtain 
\begin{equation}
\partial_\tau(\tau T^{\tau \nu}) +\partial_\eta(\tau T^{\eta \nu} )= 0, \label{eq:taueta}
\end{equation}
where $\nu= t, x, y$ or $z$. 
If the transverse components of the four-velocity $u^x$ and $u^y$ are vanishing, Eq.\:\eqref{eq:taueta} expresses the one-dimensional longitudinal expansion. 
In our numerical scheme, we utilize the Lagrange step \cite{Van1979} in which the grid-cell boundary itself moves together with the hydrodynamic flow during a time step from $\tau^n$ to $\tau^{n+1}=\tau^n+\Delta\tau$. 
We discretize Eq.\:\eqref{eq:taueta} by space-time integration in a grid cell based on the Lagrangian approach (Fig.\:\ref{fig:taueta-lat}), 
\begin{equation}
\int_{\tau^n}^{\tau^{n+1}}\int^{\eta_{i+1}+\delta\eta_{i+1}(\tau')}_{\eta_i+\delta\eta_i(\tau')}
\partial_\alpha (\tau T^{\alpha \nu}) d\eta d\tau=0, \label{eq:int}
\end{equation}
where $\alpha=\tau$ or $\eta$.
Here $\eta_i$ is the location of the $i$th grid-cell boundary at the proper time $\tau^n$, $\delta \eta_i(\tau')$ expresses a moving distance of the grid-cell boundary from $\tau=\tau^n$ to $\tau=\tau^n+\tau'$ $(0\leq \tau'\leq \Delta\tau)$, and $\eta_i + \delta \eta_i(\tau')$ indicates the location of the $i$th grid-cell boundary at the proper time $\tau$. 
The center of the $i$th grid cell is located at $\eta_i +\Delta\eta/2$ at $\tau = \tau^n$, where $\Delta\eta=\eta_i -\eta_{i-1}$.
Using Gauss' theorem for integration of Eq.\:\eqref{eq:int}, we find the value of $T^{\tau \nu}$ of the $i$th grid cell at the next time step $\tau^{n+1}$,  
\begin{align} &( T^{\tau \nu} )^{n+1}_i =\frac{\tau^n\Delta\eta}{\tau^{n+1} \Delta\eta_i^{{\rm lag}}} ( T^{\tau \nu})^n_i 
 - \frac{1}{\tau^{n+1} \Delta\eta_i^{{\rm lag}}} \nonumber\\
&\quad\;\; \times\left [\int_{C_i} T^{\alpha \nu}n_{\alpha,i} ds -\int_{C_{i-1}}T^{\alpha \nu}n_{\alpha,i-1}ds  \right],\label{eq:evolve}
\end{align}
where $\Delta\eta^{\rm lag}_i\equiv \Delta\eta +(\delta\eta_i(\Delta\tau) -\delta\eta_{i-1}(\Delta\tau))$ is the Lagrange grid-cell size at the proper time $\tau^{n+1}$, $C_i$ is the trajectory of the Lagrange grid-cell boundary at $\eta_i$ and $n_{\alpha, i}$ is the unit normal vector to $C_i$ (Fig.\:\ref{fig:taueta-lat}). 
The average values of the conserved quantities in the grid cell at the proper times $\tau^n$ and $\tau^{n+1}$ are defined respectively, by 
\begin{align}
( T^{\tau \nu})^{n}_i &\equiv \frac{1}{\Delta \eta}\int ^{\eta_{i}}_{\eta_{i-1}} T^{\tau \nu}(\tau^n,\eta) d\eta,\\
( T^{\tau \nu})^{n+1}_i &\equiv \frac{1}{\Delta \eta_i^{{\rm lag}}}\int ^{\eta_{i}+\delta\eta_i(\Delta\tau)}_{\eta_{i-1}+\delta\eta_{i-1}(\Delta\tau)} T^{\tau \nu}(\tau^{n+1},\eta) d\eta.
\end{align}
The second term of Eq.\:\eqref{eq:evolve} indicates the flux of the conserved quantities passing through the grid-cell boundary. 

\par
Using the analytical solution of the Riemann problem in Milne coordinates, we evaluate the $\Delta\eta_i^{\rm lag}$ and the flux term in Eq.\:\eqref{eq:evolve}.
The grid-cell boundary corresponds to the initial discontinuity in the Riemann problem and moves together with the contact discontinuity in the solution of the Riemann problem.
This indicates that the physical quantities on the trajectory $\delta\eta_i(\tau')$ are given by those on the contact discontinuity. 
We obtain the moving distance $\delta\eta_i(\Delta\tau)$ of the grid-cell boundary
\begin{equation} \delta\eta_i(\Delta\tau)= \frac{1}{2}{\rm log} \left( \frac{(1+V^z_{0,i})\delta t(\Delta\tau) + \tau^n}{(1-V^z_{0,i})\delta t (\Delta\tau)+ \tau^n}\right), \label{eq:deltaeta}
\end{equation}
using $\delta t(\Delta\tau)$ 
\begin{align}
&\delta t(\Delta\tau)=\frac{1}{1-(V^z_{0,i})^2}\biggl\{ -\tau^n  \nonumber \\ 
&\qquad +\sqrt{(\tau^n)^2 +(1-(V_{0,i}^z)^2) (2\tau^n\Delta\tau +{\Delta\tau}^2)}\biggr\}. 
\label{eq:deltat}
\end{align}
In Eqs.\:\eqref{eq:deltaeta} and \eqref{eq:deltat}, $V_{0.i}^z$ is the velocity of the grid-cell boundary at $\eta_i$ seen from an observer sitting at $\eta=\eta_i$ in Milne coordinates.
In the construction of algorithm, we use the Lorentz boost transformation which is explained in the next paragraph. 
Here we show the explicit form of Eq.\:\eqref{eq:evolve}. 
For detailed derivation of it please see \ref{app-flux}.
Up to the third order in $\Delta \tau$, the flux terms are given by
\begin{align} 
& \int_{C_i} T^{\alpha t}n_{\alpha,i} ds =P_i({\rm sinh}\eta_i+V^z_{0,i}{\rm cosh}\eta_i ) \Biggl\{ \Delta\tau \nonumber\\
&\;+\frac{(V^z_{0,i})^2}{2\tau^n}\Delta\tau^2+\frac{1}{2(\tau^n)^2}\left( (V^z_{0,i})^4 -(V^z_{0,i})^2\right)  \Delta\tau^3 \Biggr\}, \label{eq:flux1} \\
& \int_{C_i} T^{\alpha z}n_{\alpha,i} ds =P_i({\rm cosh}\eta_i+V^z_{0,i}{\rm sinh}\eta_i )\Biggl\{ \Delta\tau \nonumber\\
&\;+\frac{(V^z_{0,i})^2}{2\tau^n}\Delta\tau^2+\frac{1}{2(\tau^n)^2}\left( (V^z_{0,i})^4 -(V^z_{0,i})^2\right)  \Delta\tau^3 \Biggr\} , \label{eq:flux2}\\
&\int_{C_i} T^{\alpha j}n_{\alpha,i} ds = 0 ,\qquad j=x,y, 
 \label{eq:flux3}
\end{align}
where $P_i$ is the pressure on the cell boundary located at $\eta_i$.

\par From now we explain the numerical algorithm for solving the discretized equation Eq.\:\eqref{eq:evolve}. 
We express the hydrodynamic variables in Milne coordinates as $\bm W\equiv (n, w^x, w^y, w^\eta, p)$. 
The first step is the interpolation procedure in which  the 
left and right hydrodynamic states at the grid-cell boundary $\bm W_{S,i}=(n_{S,i}, w^x_{S,i},w^y_{S,i},w^\eta_{S,i}, p_{S,i})$  $(S=L,R)$ 
are determined from  the reconstruction of the distribution of volume-averaged hydrodynamic variables $\bm W$ in 
a grid cell. 
If a linear interpolation method is used for reconstruction of the distribution of hydrodynamic variables, second-order accuracy is achieved (the MC limiter \cite{Van1979}). 
For third-order accuracy, we need to utilize a quadratic curve (the piecewise parabolic method (PPM) \cite{Mignone2005, Colella1984, Marti1996}) in reconstruction of the distribution of hydrodynamic states. 
For the test calculation in the next section, we use the PPM.
Next, using the constructed left and right states $\bm W_{S,i}$ in Milne coordinates, 
we prepare the initial condition of the Riemann problem in terms of $\bm V_{S,i}$ in Cartesian coordinates. 
To obtain $\bm V_{S,i}$, 
we move a grid cell to $\eta=0$ with the Lorentz boost transformation and perform the coordinate transformation to $\bm V$. 
Here $\bm W$ is invariant under the Lorentz boost transformation. 
Using these relations, $v^{x,y}=w^{x,y}$ and $v^z = \tau w^\eta$ at  $\eta=0$, 
we obtain 
\begin{equation} 
\bm V_{S,i}= (n_{S,i}, w^x_{S,i},w^y_{S,i},\tau w^\eta_{S,i}, p_{S,i}) \label{eq:Vs}.  
\end{equation}

\par In the second step, we solve the Riemann problem 
Eq.\:\eqref{eq:riemann} with the initial condition Eq.\:\eqref{eq:Vs}. 
 $P_i$ and $V^z_{0,i}$ in Eqs.\:\eqref{eq:deltaeta}-\eqref{eq:flux3} 
are determined by the analytical solution of the Riemann problem.\footnote{At the boundary between the matter and  the vacuum in Sect.\:\ref{subsub:vacuum}
we use the values of the vacuum, $V^z_0=1, P=0$, instead of solving the Riemann problem.}
For solving the Riemann problem we employ the two-shock approximation \cite{Colella1982,Balsara1994,Dai1997,Mignone2005}. In the approximation, we can avoid solving the ordinary differential equation for the rarefaction wave which takes a lot of computational time in the multidimensional problem. 
The numerical procedure of this step depends on the equation of state. The Riemann solution with the  QCD equation of state 
is described in Ref. \cite{Akamatsu2014}.
After solving the Riemann problem, we perform the inverse Lorentz transformation to the original frame. 
Here $\Delta\eta^{\rm lag}_i$ is Lorentz invariant. 
The flux terms Eqs.\:\eqref{eq:flux1}-\eqref{eq:flux3} are defined on the original frame but written in terms of $V^z_{0,i}$. 

\par In the third step, solving the discretized hydrodynamic equation Eqs.\:\eqref{eq:evolve}, \eqref{eq:deltaeta}-\eqref{eq:flux3} with the values of $P_i$ 
and $V_{0,i}^z$,
we obtain distribution of the conservative quantities $T^{\tau \nu}(\nu =t,x,y,z)$ at the next time step using the Lagrange scheme. 
We remap the grids which move in the Lagrange step on the Eulerian coordinate \cite{Van1979}.
\footnote{We do not perform the remap step at the boundary between matter and the vacuum to keep the exact location of  the boundary determined by the Lagrange grid.}

\par  In the final step, we construct the primitive variables $\bm W$ from the conserved quantities \cite{Akamatsu2014}. 
We obtain $T^{\tau\alpha}(\alpha = \tau,x,y,\eta)$ from $T^{\tau \mu}(\mu=t,x,y,z)$ from the coordinate transformation. 
The numerical method for construction of $\bm W$ from $T^{\tau\alpha}$ is the same as that in Ref.\cite{Akamatsu2014}. 

\subsubsection{(3+1) dimensional systems} \label{sec:3d}
The one-dimensional code is easily extended to a multidimensional code by the Strang splitting method \cite{Strang1968}, that is to say, multidimensional hydrodynamic evolution is realized by successive one-dimensional hydrodynamic calculations. 
To avoid counting the expansion effect of coordinates more than once, we extract one-dimensional hydrodynamic equations from Eqs.\:\eqref{eq:source1}-\eqref{eq:source4}, which have source terms in the right-hand side. 
Then we rewrite each one-dimensional hydrodynamic equation as that without the source terms.  

To be explicit, we express Eqs.\:\eqref{eq:source2}-\eqref{eq:source4} as 
\begin{equation} 
\partial_\tau T^{\tau \alpha} + \partial_i T^{i \alpha} + \partial_\eta T^{\eta \alpha}=S^\alpha, 
\label{eq:source-all}
\end{equation}
where the source term $S^\alpha$ is given by 
\begin{equation} 
S^\alpha =\left( -\frac{1}{\tau}T^{\tau\tau} -\tau T^{\eta\eta},\: - \frac{1 }{\tau}T^{\tau x},\: - \frac{1}{\tau}T^{\tau y}, \:-\frac{3}{\tau}T^{\tau\eta} \right). 
\end{equation}
Applying the dimensional splitting method to Eq.\:\eqref{eq:source-all}, we obtain the one-dimensional hydrodynamic equations, 
\begin{align} 
&\partial_\tau T^{\tau \alpha}+\partial_x T^{x\alpha} =0, 
\label{eq:split-x} \\
&\partial_\tau T^{\tau \alpha}+\partial_y T^{y\alpha} =0, 
\label{eq:split-y} \\
&\partial_\tau T^{\tau \alpha}+\partial_\eta T^{\eta \alpha}=S^\alpha. 
\label{eq:split-eta}
\end{align}
Solving the one-dimensional equations Eqs.\:\eqref{eq:split-x}-\eqref{eq:split-eta} successively, we carry out the (3+1) dimensional calculations.

For expansion in the rapidity direction, transforming Eq.\:\eqref{eq:split-eta} into Eq.\:\eqref{eq:taueta}, we use the algorithm mentioned 
in Sect.\:\ref{sec:1d}. 
In the $x$ and $y$ directions, we use the same algorithm in Cartesian coordinates in Ref.\cite{Akamatsu2014}. 
For example, for expansion in the $x$ direction, we discretize Eq.\:\eqref{eq:split-x} as explained in Sec.\ref{sec:1d}, 
\begin{align}
 &( T^{\tau \tau} )^{n+1}_i =\frac{\Delta x}{\Delta x_i^{{\rm lag}}} ( T^{\tau \tau})^n_i \nonumber\\
& \qquad\qquad\quad - \frac{\Delta\tau}{\Delta x_i^{{\rm lag}}}\left (P_iV^x_{0,i} -P_{i-1}V^x_{0,i-1} \right),\label{eq:evolvex1}\\ 
& ( T^{\tau x} )^{n+1}_i =\frac{\Delta x}{\Delta x_i^{{\rm lag}}} ( T^{\tau x})^n_i 
- \frac{\Delta\tau}{\Delta x_i^{{\rm lag}}}\left (P_i -P_{i-1} \right),\label{eq:evolvex2} \\
&( T^{\tau y} )^{n+1}_i =\frac{\Delta x}{\Delta x_i^{{\rm lag}}} ( T^{\tau y})^n_i,  \label{eq:evolvex3}
\end{align}
where $P_i$ and $V^x_{0,i}$ are determined by the solution of the Riemann problem whose initial condition is given by Eq.\:\eqref{eq:Vs}. 
The grid-cell size after the Lagrange step is given by 
\begin{equation} \Delta x^{\rm lag}_i=\Delta x + (V^x_{0,i} - V^x_{0,i-1}) \Delta\tau. 
\end{equation}

Using the operator $L^k_i$ which represents one-dimensional evolution in the $i$ direction during the proper time $k\Delta \tau$, 
two-dimensional expansion in $(x,\eta)$ coordinates is given by $L^k_x$ and $L^k_\eta$, 
\begin{equation} 
\bm (T^{\tau\alpha})^{n+1}=L^{1/2}_x L_\eta^1 L_x^{1/2} \bm (T^{\tau\alpha})^n. 
\end{equation}
Similarly the three-dimensional expansion in $(x,y,\eta)$ coordinates  is written by 
\begin{align} 
\bm (T^{\tau\alpha})^{n+1} =&L^{1/6}_x L^{1/6}_y L_\eta^{1/3}L^{1/6}_y L_x^{1/3} L_\eta^{1/6} L_y^{1/3} L_x^{1/6} \nonumber\\
&\times L_\eta^{1/3} L_x^{1/6} L_y^{1/3} L_\eta^{1/6} L_x^{1/6} \bm (T^{\tau\alpha})^n.
\end{align}

\subsection{The Courant-Friedrichs-Lewy condition}\label{sec:CFL}
The Courant-Friedrichs-Lewy (CFL) condition helps us determine an appropriate  time-step size in solving partial differential equations. 
The CFL condition is determined so that the numerical propagating speed on the grid is larger than the physical propagating speed,
\begin{equation} 
\frac{\tau\Delta \eta}{\Delta \tau} > \tau |w^\eta_c| \label{eq:CFL}, 
\end{equation}
where $\Delta \eta/\Delta \tau$ and $w_c^\eta$ are the numerical signal velocity and the characteristic velocity in the rapidity direction, respectively.
Note that the proper time $\tau$ is multiplied for dimensionless expression.
This condition is important for the stability of the numerical calculations.
If the CFL condition Eq.\:\eqref{eq:CFL} is written as a function of the characteristic velocity in $z$ direction $v^z_c$,

\begin{equation}
\frac{\tau\Delta\eta}{\Delta\tau}  > \left|\frac{-{\rm sinh}\eta - |v^z_c|{\rm cosh}\eta}{{\rm cosh}\eta + |v^z_c| {\rm sinh}\eta}\right| \qquad {\rm for } \;\; \eta >0, \label{eq:cfl1} 
\end{equation}

\begin{equation}
\frac{\tau\Delta\eta}{\Delta\tau}  > \left|\frac{-{\rm sinh}\eta +|v^z_c|{\rm cosh}\eta}{{\rm cosh}\eta - |v^z_c| {\rm sinh}\eta}\right| \qquad {\rm for }\;\; \eta <0, \label{eq:cfl2}
\end{equation}

the CFL condition in Milne coordinates has a rapidity dependence. 
At large rapidity the right-hand side of Eqs.\:\eqref{eq:cfl1} and \eqref{eq:cfl2} approaches 1, which is the same condition as that in the case of $|v^z_c|=1$.
Therefore the CFL condition in Milne coordinates is determined by 
\begin{equation} 
\frac{\Delta\tau}{\tau \Delta\eta}  < 1.  \label{eq:CFLtau}
\end{equation}
Equation \eqref{eq:CFLtau} indicates that the proper time of the denominator is larger, a possible time-step size $\Delta \tau$ is larger. 
With these considerations, we define the Courant number as 
\begin{equation}
C\equiv \frac{\Delta \tau}{\tau_0\Delta\eta}\quad (0<C<1),  \label{eq:Courant}
\end{equation}
with an initial proper time $\tau_0$.

\begin{figure*}[t!]
  \begin{minipage}{0.5\hsize}
  \centering
  \includegraphics[width=8cm]{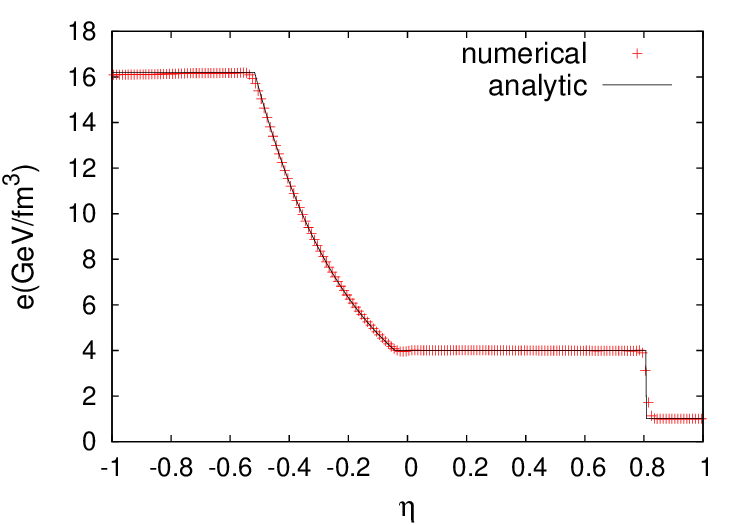}
  \end{minipage}
  \begin{minipage}{0.49\hsize}
  \centering
  \includegraphics[width=8cm]{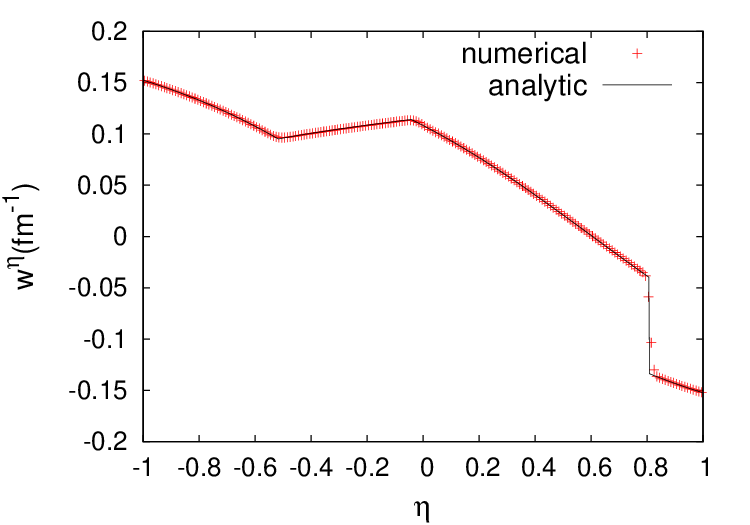}
  \end{minipage}
  \centering
  \includegraphics[width=8cm]{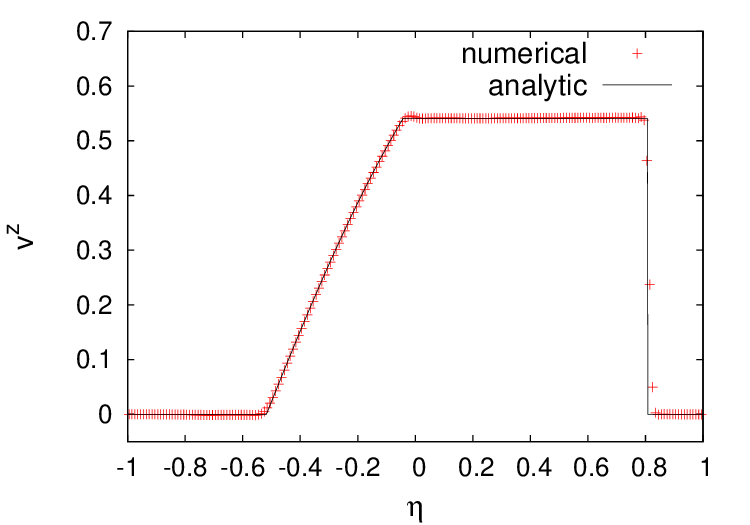}
  \caption{Comparison between the analytical solution (solid line) and the numerical calculations for the shock tube problem at $\tau = 5$ fm. 
   The initial discontinuity is located at  $\eta=0$. 
   Top left : The energy density distribution as a function of $\eta$. 
   Top right: The velocity distribution $w^\eta$ as a function of $\eta$. 
   Bottom: The velocity distribution $v^z$ as a function of $\eta$. }
  \label{fig:shocktube1}
\end{figure*}

\begin{figure*}[t!]
  \begin{minipage}{0.5\hsize}
  \centering
  \includegraphics[width=8cm]{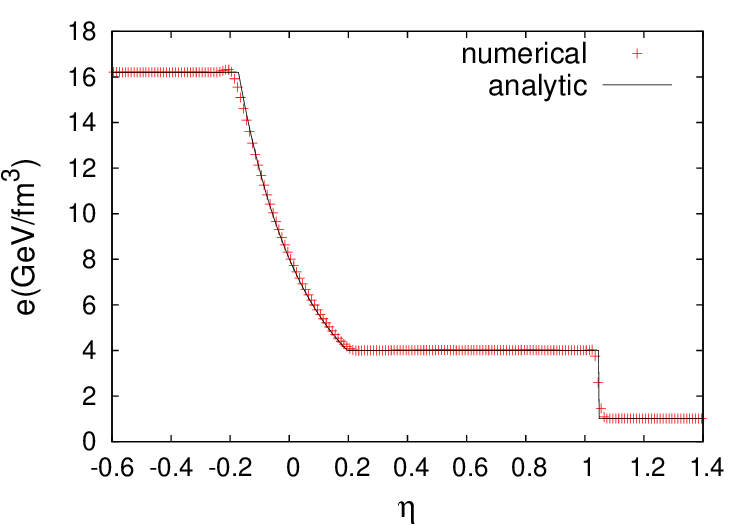}
  \end{minipage}
  \begin{minipage}{0.49\hsize}
  \centering
  \includegraphics[width=8cm]{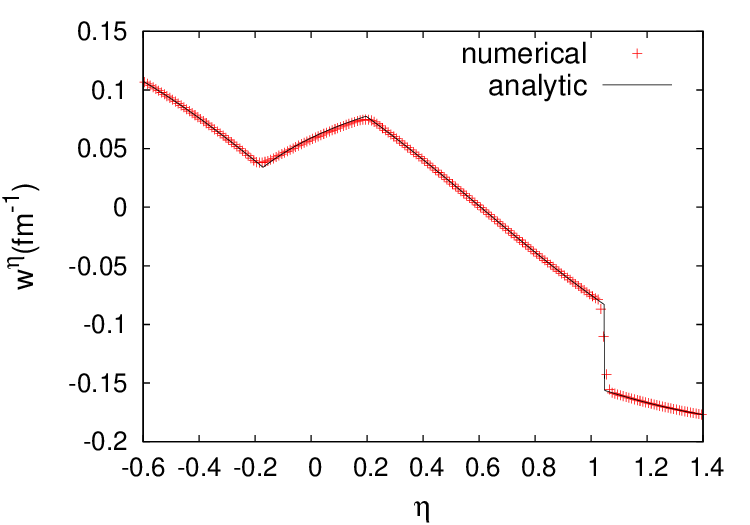}
\end{minipage}
\centering
  \includegraphics[width=8cm]{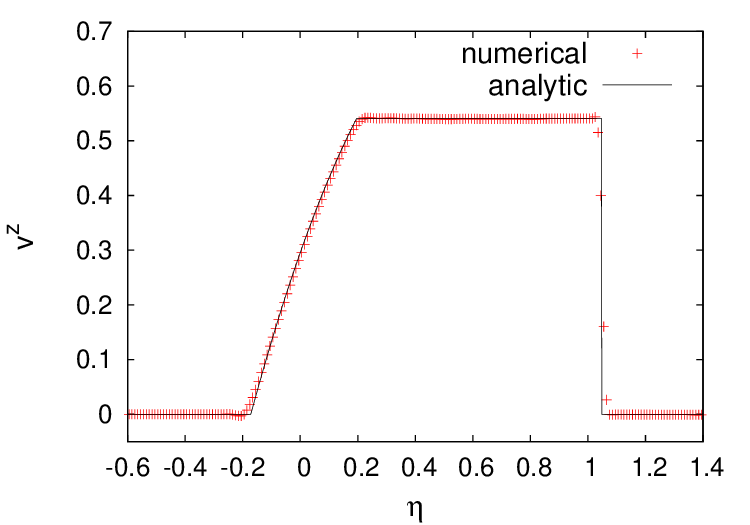}
 \caption{Comparison between the analytical solution (solid line) and the numerical calculations for the shock tube problem at $\tau = 5$ fm. The initial discontinuity is located at  $\eta=1$. 
 Top left: The energy density distribution as a function of $\eta$. 
 Top right: The velocity distribution $w^\eta$ as a function of $\eta$. 
 Bottom: The velocity distribution $v^z$ as a function of $\eta$.}
  \label{fig:shocktube2}
\end{figure*}

\subsection{Boundary conditions}
When we numerically solve partial differential equations and update the value of a cell, the values of its neighboring cells are necessary.
For the cell on the boundary of hydrodynamic grid, we need to prepare additional cells which are called ghost cells and put appropriate information to them. 
In the next section we shall discuss several numerical tests using our code; shock tube problems in Milne coordinates, expansion of matter  to the vacuum, Landau-Khalatnikov solution, fluctuations in longitudinal expansion, Gubser flow, and the conservation property. 
For the shock tube problem in Milne coordinates, we input an analytical solution as a boundary condition at the ghost cells.
For the numerical test of expansion of matter to the vacuum, we use the physical values of the vacuum at the cell on the boundary. 
We employ the periodic boundary condition for the investigation of fluctuations in longitudinal expansion. 
For the Landau-Khalatnikov solution and the Gubser flow, we copy the values of the cells on the boundary onto those of the ghost cells.

In addition, we point out that in some cases we need a careful procedure at the boundary. 
For example, for the shock tube problem in Milne coordinates, we use the MC limiter at the boundary and ghost cells to 
reduce numerical errors which originate from inward flow at the boundary. 
In the expansion of matter to the vacuum, we observe that a numerical instability occurs at the discontinuity between matter and the vacuum. 
To stabilize the difficulty we employ the minmod limiter which is dissipative and smears out discontinuities compared to the MC limiter. 


\section{Numerical tests}\label{sec:numerical_test}

We employ several problems to check correctness of the numerical algorithm in Milne coordinates. 
For one-dimensional tests, we analyze with our code the Riemann problem and Landau--Khalatnikov solution \cite{Landau1953, Landau1956} in order to verify our Riemann solver.
The Landau-Khalatnikov solution is used for understanding the experimental data of the particle rapidity distributions in the high-energy heavy-ion collisions. 
Next we discuss propagation of fluctuations around Bjorken flow. 
We derive analytical solutions from linearized hydrodynamics and compare them to numerical calculations with our code. 
For multidimensional tests, we use the Gubser flow \cite{Gubser2010, Gubser2011}, which gives us a three-dimensional hydrodynamic expansion of hot and dense matter created after the high-energy heavy-ion collisions. 
In the test problems we use the ideal gas equation of state, $e=3p$.

\subsection{Riemann problem}
\subsubsection{Shock tubes \label{sec:shoch-tube}}

\begin{figure}[b]
  \centering
  \includegraphics[width=8cm]{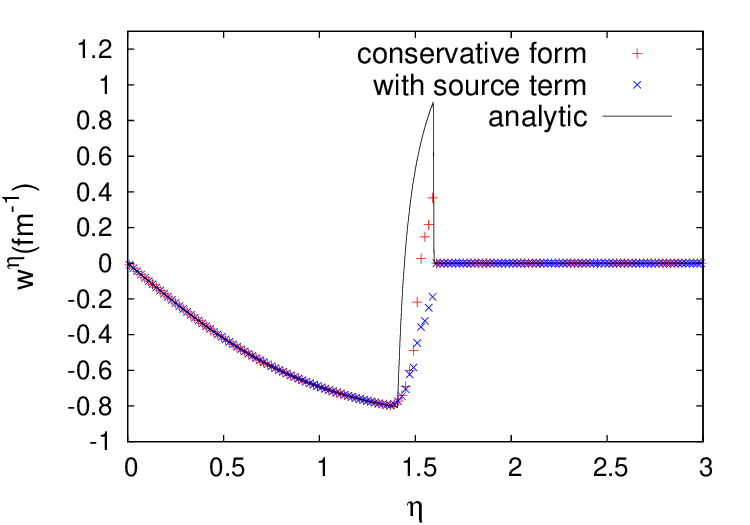}
  \caption{The numerical results of codes with and without the source terms and the analytical solution (the solid line) of $w_\eta$ as a function of $\eta$ for the expansion into vacuum at $\tau=1.1$ fm. }
   \label{fig:vac_t11}
\end{figure}
\begin{figure}[ht]
  \centering
  \includegraphics[width=8cm]{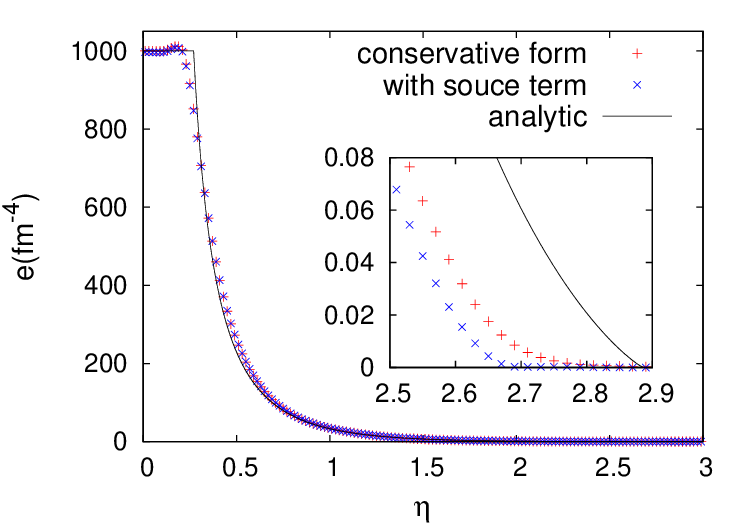}
  \centering
  \includegraphics[width=8cm]{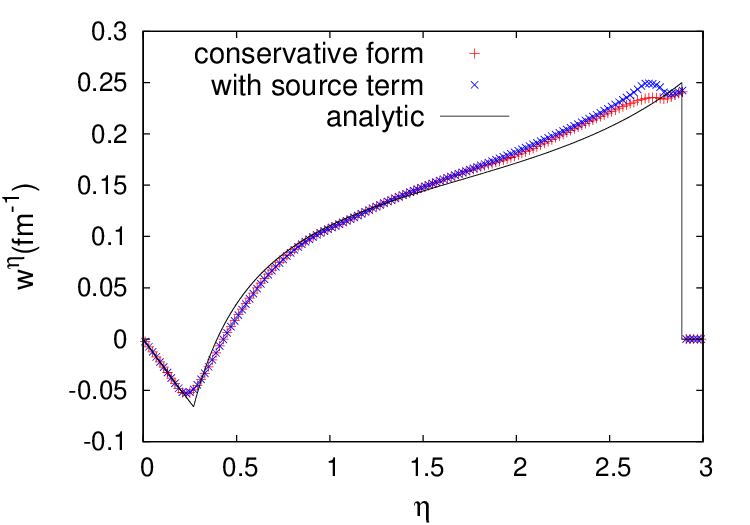}
\centering
  \includegraphics[width=8cm]{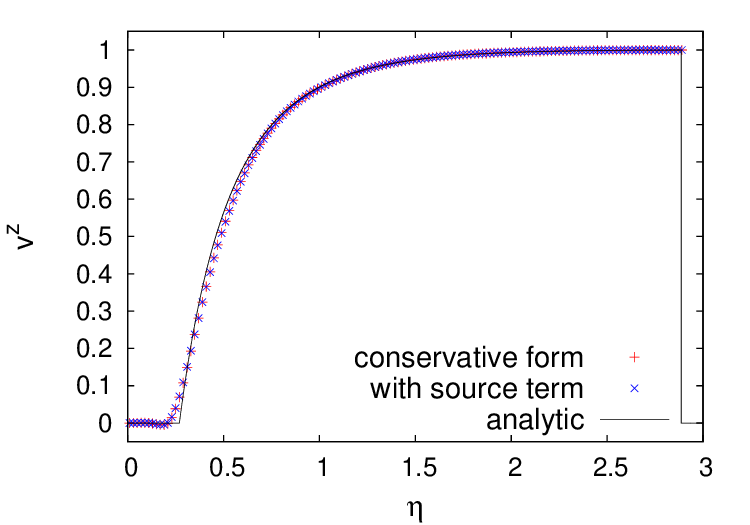}
 \caption{The numerical results of codes with and without the source terms of the energy density (top left), $w^\eta$ (top right) and $v^z$ (bottom) distributions as a function of $\eta$ together with the analytical solution (solid line) for the expansion into vacuum at $\tau=4$ fm.}
  \label{fig:vacuum}
\end{figure}

Using the property that the initial-value problem of Eq.\:\eqref{eq:initial} in Milne coordinates is the same as the Riemann problem in Cartesian coordinates,
we carry out the shock tube test in Milne coordinates. 
We solve the initial-value problem of Eq.\:\eqref{eq:initial} with our numerical scheme in Milne coordinates and compare the numerical results with the analytical solution of the Riemann problem Eq.\:\eqref{eq:riemann}.
The analytical solution of the Riemann problem in Milne coordinates is obtained by the coordinate transformation from that in Cartesian coordinates \cite{Marti1994, Pons2000} (see \ref{app-Riemann-taueta}). 

\par  We perform test calculations in two cases. In the first case we set the discontinuity at $\eta=0$ and in the second case we put it at $\eta=1$ at the initial time $\tau_0=1$ fm. 
In the first test problem,  the temperature in $\eta>0$  ($\eta<0$) is set  to $T_L=400\:{\rm MeV}$ ($T_R=200\:{\rm MeV}$) and in both regions the $z$ component of the velocity $v^z$ is vanishing.   
In our numerical scheme, we directly calculate the time evolution of the rapidity component of the velocity $w^\eta$, instead of $v^z$.  
The transformation between $v^z$ and $w^\eta$ is given by Eq.\:\eqref{eq:trans-veta}, which suggests that $w^\eta$ is not vanishing even if $v^z=0$. 
The initial condition of $w^\eta$ is given by $w^\eta=\frac{1}{\tau_0}\frac{-{\rm sinh}\eta }{{\rm cosh}\eta }$ which has an $\eta$ dependence.
We perform the numerical calculations with grid size $\Delta\eta=0.01$ and time-step size $\Delta\tau=0.1 \tau_0\Delta\eta$.  
Figure \ref{fig:shocktube1} shows the energy density distribution, the velocity $w^\eta$, and  the velocity $v^z$, which is 
transformed by Eq.\:\eqref{eq:trans-veta} from $w^\eta$ as a function of rapidity $\eta$ together with the analytical solutions (solid lines).   
The rarefaction wave moves to the negative direction from $\eta=0$ and the shock wave moves to the positive direction from $\eta=0$. 
Our numerical results are consistent with the analytical solutions.

\par For the second numerical test, we put the initial discontinuity at $\eta\not=0$. 
Other conditions, $T_L$, $T_R$ and $v^z$ are the same as those in the first test calculation. 
In Cartesian coordinates the first and second numerical tests are essentially identical. 
In Milne coordinates, however, the second numerical problem is different from the first problem, because the $w^\eta$ depends on $\eta$ differently from the first problem. 
The energy density distribution $e$, the velocity distribution $w^\eta$, and the velocity distribution $v^z$ are shown in Fig.\:\ref{fig:shocktube2}. 
Again, our numerical calculations show good agreement with the analytical solutions in the second numerical test.

\subsubsection{Expansion of matter into the vacuum} \label{subsub:vacuum}
As one of the specific problems of the Riemann problem, we consider the one-dimensional expansion of matter into the vacuum;  
a rarefaction wave appears at the discontinuity and expands between the matter and the vacuum \cite{Rischke1995}. 
This problem is useful for a realistic description of expansion of the QGP and hadronic matter into the vacuum in the high-energy heavy-ion collisions.  
We set the initial condition to 
\begin{equation}
\begin{array}{lll} p=1000 \;{\rm fm}^{-4}, & \quad w^\eta= -\frac{{\rm sinh}\eta}{{\rm cosh}\eta} &\quad {\rm for}\quad |\eta| \leq 1.5, \\
 p=0, &\quad w^\eta=0, &\quad {\rm for} \quad |\eta|>1.5, 
 \label{eq:vacuum}
\end{array} \end{equation}
where $w^\eta$ in $|\eta| \leq 1.5$ corresponds to $v^z=0$  in Cartesian coordinates and  in $|\eta| > 1.5$ lies the vacuum. 
Setting the vacuum for the boundary condition as in Eq.\:\eqref{eq:vacuum}, one can avoid the matter from flowing into the system through the boundaries. 
In the shock tube problems in Milne coordinates in Sect.\ref{sec:shoch-tube}, the matter comes in through the boundaries, which is 
a possible source of numerical error. 

Here we discuss the importance of description of the hydrodynamic equation in the conservative form in developing numerical algorithm. 
For the investigation, 
we discretize the hydrodynamic equation with the source terms Eq.\:\eqref{eq:split-eta} and 
construct a code based on the same procedure explained in Sect.\:\ref{sec:1d}. 
In the code, $T^{\tau\tau}$ and $T^{\tau\eta}$ are updated in the Lagrange step, instead of $T^{\tau t}$ and $T^{\tau z}$.
Then we compare the results of the code without the source terms, Eq.\:\eqref{eq:taueta}, and those with the source terms.
In both numerical calculations the grid size $\Delta\eta$ and the time-step size are set to 0.02  and $0.1\tau_0 \Delta\eta$, 
respectively. 
The hydrodynamic expansion starts at $\tau_0=1$ fm.

Figure \:\ref{fig:vac_t11} shows the $w^\eta$ distribution as a function of $\eta$ at $\tau=1.1$ fm. 
In $0<\eta<1.46$, the matter is at rest in Cartesian coordinates, which corresponds to the negative flow in Milne coordinates. 
In $1.46<\eta<1.58$, the rarefaction wave starts to expand; it propagates with the sound velocity inward the matter at rest in Cartesian coordinates and expands with the speed of light to the vacuum.  
In the rarefaction wave the steep velocity gradient is produced, where both of the codes with and without the source terms cannot reproduce the analytical result, though the code without the source terms is closer to the analytical solution.

Figure \:\ref{fig:vacuum} shows the numerical results of the energy density, $w^\eta$  and $v^z$ distributions as a function of $\eta$ together with the analytical solution at later time $\tau=4$ fm.
In $|\eta|<0.27$ the velocity $v^z$ is vanishing in Cartesian coordinates, which indicates the negative $w^\eta$ in Milne coordinates. 
In $0.27<|\eta|<2.89$ the rarefaction wave is spreading and at $\eta=\pm 2.89$ the boundary to the vacuum exits, which moves out to the vacuum at the speed of light. 
For the energy density distribution, both codes with and without the source terms reproduce the analytical solution, though near the boundary between the matter and the vacuum a small difference between them is observed. 
In both codes, the value of $w^\eta$ around the boundary is larger than that of the analytical solution near the boundary. 
The stronger flow near the boundary in the numerical solution causes the smaller energy density compared with the analytical solution. 
On the other hand, there is no difference between the behaviors of $v^z$ of both codes. 
The stability of numerical calculation is sensitive to the differences in $w^\eta$ of codes 
which are seen near the boundary between matter and the vacuum, because at the boundary pressure 
becomes zero and the Lorentz factor becomes infinity in the analytical solution.

\begin{figure}[t]
  \centering
  \includegraphics[width=8cm]{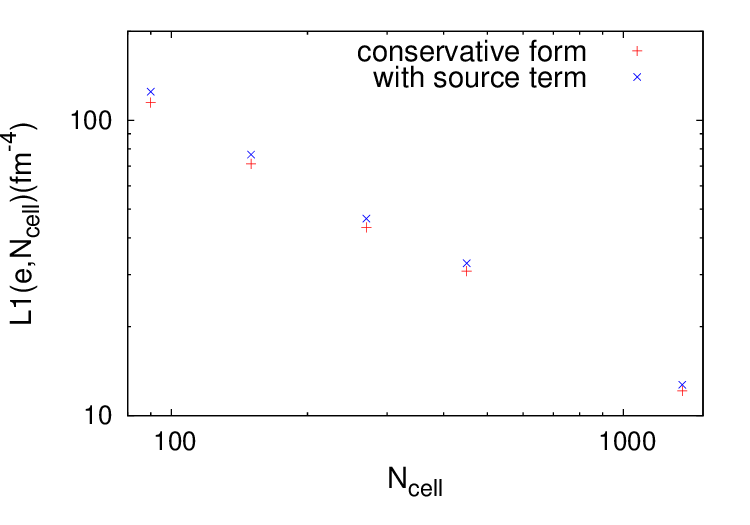}
  \centering
  \includegraphics[width=8cm]{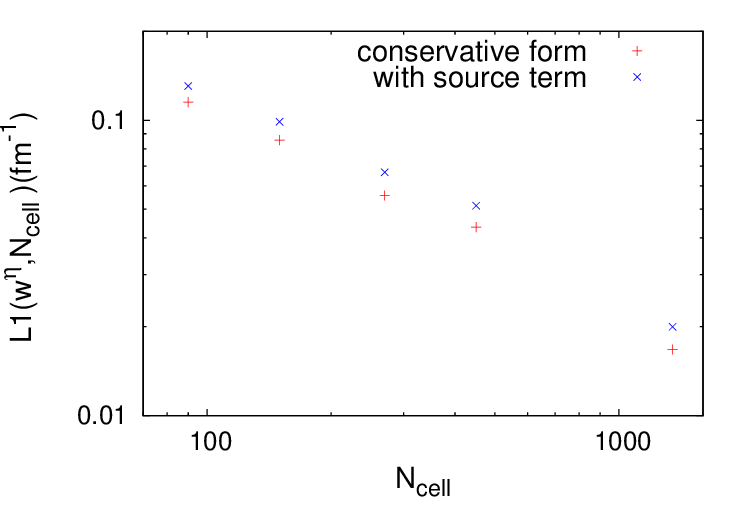}
 \caption{
 The L1 norm of the energy density $e$ (top panel) and the velocity $w^\eta$ (bottom panel) for the expansion into vacuum
 as a function of number of cell for the codes with and without the source terms.}
  \label{fig:vacL1}
\end{figure}

To investigate the numerical accuracy of the codes with and without the source terms, we calculate the L1 norm which is defined by 
\begin{equation} 
L(u,N_{\rm cell})=\sum_{i} \left| u(\eta_i;N_{\rm cell}) -u_{\rm exact}(\eta_i)\right| \Delta \eta,
\label{eq:L1-v}
\end{equation}
where $u$ is the energy density $e$ or the rapidity component of the velocity $w^\eta$, $\Delta\eta$ is a grid-cell size. 
Using Eq.\:(\ref{eq:L1-v}), we evaluate the deviation of the numerical results $u(\eta_i ; N_{\rm cell})$ from the exact solutions.
At the same time we can know the convergence speed to the exact solution of  numerical algorithm. 
In Fig.\:\ref{fig:vacL1}, the L1 norms of energy density and $w^\eta$ are shown. 
In both energy density and velocity, the values of the L1 norm of the code without the source terms are smaller than those with the source terms, which means that the code without  the source term has smaller numerical viscosity than that with the source terms. 
As expected, the existence of the source terms produces more artificial viscosity. 
 
If the initial discontinuity is set at the larger rapidity, this makes the velocity slope at rarefaction wave larger and gives more severe problems to the codes in Milne coordinates. 
We find that the code without the source terms is more stable than that with the source terms. 
For example, if we set the  initial discontinuity at $\eta=1.7$, we find that numerical instability occurs in the code with the source terms.

\subsection{Landau-Khalatnikov solution}
We employ the Landau-Khalatnikov solution \cite{Landau1953, Landau1956} as a one-dimensional numerical test problem. 
The initial condition of it is expressed by a thin slab of hot and dense matter created after the collisions, which is the same as the problem 
discussed in Sect.\:\ref{subsub:vacuum}. 
In the expansion of the slab of matter, two rarefaction waves travel into the slab from both sides and start to overlap at the center of the slab. 
The region where rarefaction waves overlap is described by the Landau-Khalatnikov solution.
The asymptotic form of the Landau-Khalatnikov solution for sufficiently later time $\tau \gg \Delta$ is written by
\begin{equation}
e=e_0{\rm exp}\left\{-\frac{4}{3}\left[ 2{\rm ln}\left(\frac{\tau}{\Delta}\right) -\sqrt{{\rm ln}\left(\frac{\tau}{\Delta}\right)^2-\eta^2}\;\right] \right\}  \label{eq:landau} 
\end{equation}
and $w^\eta =0$, where $\Delta$ is the thickness of the slab. 
The asymptotic solution Eq.\:\eqref{eq:landau} is used for an investigation of the rapidity distributions of the produced particles at RHIC\:\cite{Murray2004, Bearden2005, Murray2008, Steinberg2005,Wong2008,Jiang2013}. 

\begin{figure}[b]
  \centering
  \includegraphics[width=8cm]{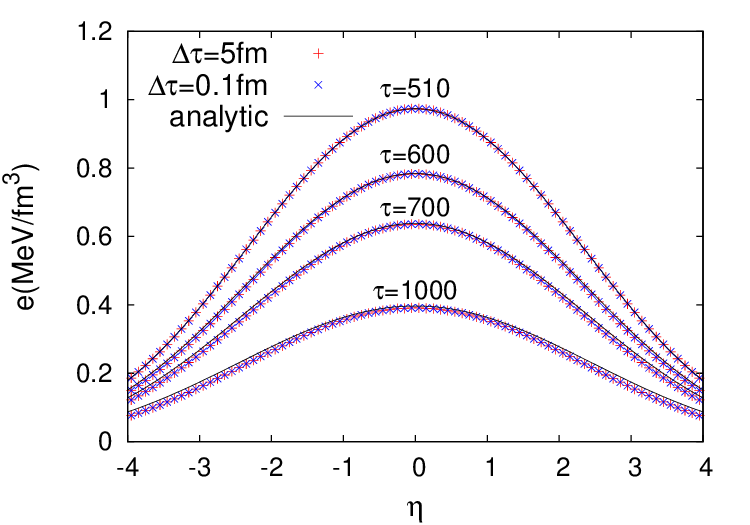}
  \caption{The analytical and numerical results of the energy density distribution for the Landau-Khalatnikov solution with $\Delta \tau =$ 0.1and 5 fm. 
  From the top, the energy density distributions at $\tau=510, 600, 700$ and 1000 fm.}
   \label{fig:landau}
\end{figure}

\par In the numerical calculation, we start the simulation at $\tau_0=500$ fm with the initial condition given by Eq.\:\eqref{eq:landau}, where $e_0$ and 
the thickness size  are set to 10 GeV/fm$^3$ and $\Delta= 0.5$ fm, respectively. 
The numerical calculation is performed with the grid size $\Delta\eta=0.1$, the time-step size $\Delta\tau=0.1$ fm 
and $\Delta\tau=0.1 \tau_0\Delta\eta=5$ fm, which is determined by the CFL condition in Sect.\:\ref{sec:CFL}.
Figure \ref{fig:landau} shows the energy density distributions at $\tau=$ 510, 600, 700, and 1000 fm together with the analytical solution. 
Calculations with the time-step sizes $\Delta\tau=5$ fm and $0.1$ fm can explain the analytical solution, which suggests that the computational
 time can be saved if the time-step size is determined by the CFL condition in Sect.\ref{sec:CFL}. 
There is a small deviation between numerical calculations and the analytical solution at large $|\eta|$, 
which implies that the asymptotic form of the Landau-Khalatnikov solution Eq.\:\eqref{eq:landau} cannot be applicable at large rapidity \cite{Landau1956}. 

\begin{figure*}[ht!]
  \begin{minipage}{0.5\hsize}
  \centering
  \includegraphics[width=8cm]{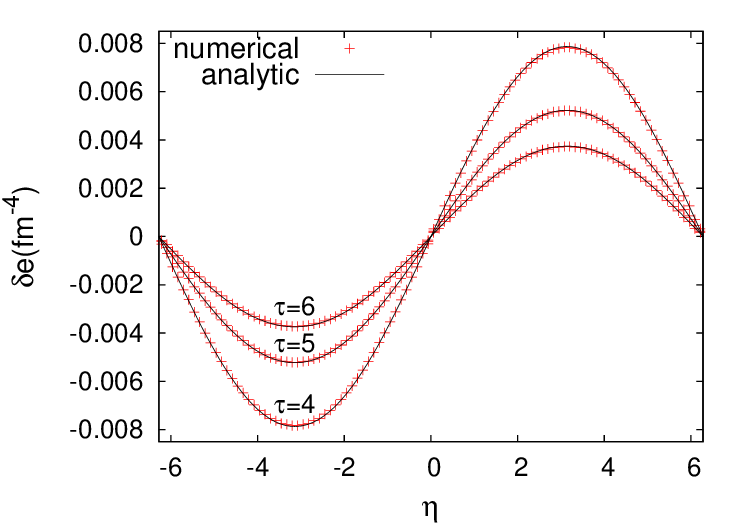}
  \end{minipage}
  \begin{minipage}{0.49\hsize}
  \centering
  \includegraphics[width=8cm]{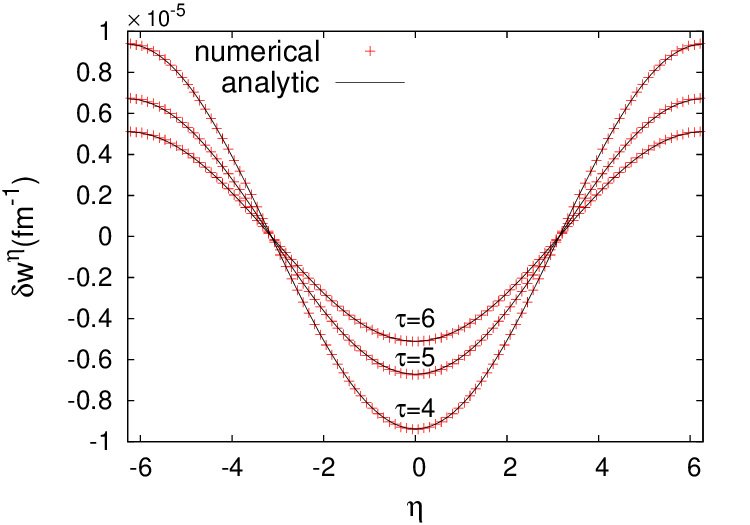}
  \end{minipage}
  \caption{Analytical and numerical results for the propagation of fluctuations around Bjorken flow at $\tau = 4$, 5 and 6 fm ($D > 0$).
  The numerical calculation is done with the grid size $\Delta x=0.1256$ fm and the time-step size $\Delta\tau = 0.1\tau_0\Delta\eta$.
  \label{fig:sounddamp}}
 \end{figure*}
\begin{figure*} [ht!]
  \begin{minipage}{0.5\hsize}
  \centering
  \includegraphics[width=8cm]{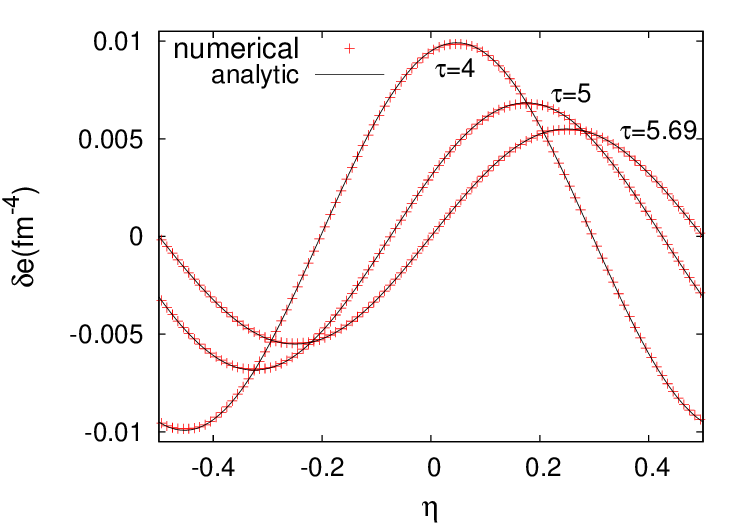}
  \end{minipage}
  \begin{minipage}{0.49\hsize}
  \centering
  \includegraphics[width=8cm]{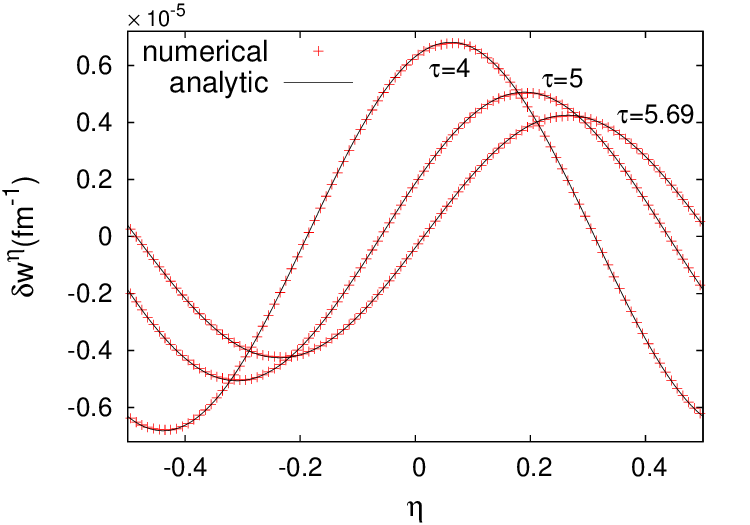}
  \end{minipage}
  \caption{Analytical and numerical results for the propagation of fluctuations around Bjorken flow at  $\tau=  4$, 5 and 5.694 fm($D<0$).
  The numerical calculation is carried out with the grid size $\Delta x=0.01$ fm and the time-step size $\Delta\tau = 0.1\tau_0\Delta\eta$.
  \label{fig:sound}}
\end{figure*}

\subsection{Propagation of longitudinal fluctuations around Bjorken flow}

The longitudinal fluctuation in particle distributions and collective flows is one of the interesting topics in high-energy heavy-ion collisions at 
LHC\:\cite{CMS2015}.
For instance, the propagation of fluctuations around Bjorken flow in heavy-ion collisions are investigated from linear analyses \cite{Baier:2006gy, Floerchinger2011}. 
The small longitudinal fluctuations around Bjorken flow propagate according to the following linearized equations:
 
\begin{align}
&e=e_B+\delta e, \quad w^\eta =\delta w^\eta,\\
&\partial_\tau \delta e+ (1+\lambda)e_B \partial_\eta \delta w^\eta+\frac{1+\lambda}{\tau}\delta e =0,  \\
& \partial_\tau \delta w^\eta +\frac{\lambda}{1+\lambda}  \frac{1}{\tau^2 e_B}\partial_\eta \delta e+\frac{2-\lambda}{\tau} \delta w^\eta =0. 
\label{Eq:Euler-fluc}
\end{align}
Here $\quad e_B=e_0\left(\frac{\tau_0}{\tau}\right)^{1+\lambda}$ is the energy density from Bjorken's scaling solution \cite{Bjorken1983} for the equation of state $p=\lambda e$ 
\footnote{In comparison between the analytical solutions and the numerical calculation, we set $\lambda$ to 1/3.}. 

Since the background is rapidity independent, we can obtain solutions with a definite wave number $\delta e, \delta w^\eta\propto e^{ik\eta}$.
The solutions consist of two modes, as Eq.\:\eqref{Eq:Euler-fluc} is essentially a second-order ordinary differential equation in the $k$-space.
The nature of the modes depends on the sign of $D \equiv (1-\lambda)^2-4k^2\lambda \neq 0$.
For $D>0$ the fluctuations do not propagate but just attenuate, while for $D<0$ the fluctuations propagate as well as attenuate.
The condition for $D=0$ is satisfied by $k = (1- c_s^2)/2 c_s$, where $c_s = \sqrt{\lambda}$ is the sound velocity. 
General solutions of the linearized equations including the case $D=0$ are given in \ref{app-general-solution}.

Here we compare the analytical solutions and the numerical calculation with our hydrodynamic code for two cases, $D>0$ and $D<0$.
In both cases, we choose initial conditions so that we can single out a particular mode of attenuation ($D>0$) and propagation ($D<0$).
To be specific, for $D>0$ we choose
\begin{align}
\delta e(\tau,\eta) =& A\left(\frac{\tau}{\tau_0}\right)^{(-3-\lambda-\sqrt{D})/2} {\rm sin}(k\eta), \\
\delta w^\eta(\tau,\eta) =& \frac{\lambda - 1-\sqrt{D}}{2ke_0(1+\lambda)\tau_0 } A \left(\frac{\tau}{\tau_0}\right)^{(-3+\lambda -\sqrt{D})/2 } {\rm cos}(k\eta), 
\end{align}
and for $D<0$
\begin{align}
\delta e(\tau,\eta) =& A\left(\frac{\tau}{\tau_0}\right)^{-(3+\lambda)/2} {\rm sin}(k\eta-\theta), \\
\delta w^\eta(\tau,\eta) &= \frac{A}{2ke_0(1+\lambda) \tau_0}\left(\frac{\tau}{\tau_0}\right)^{(\lambda-3)/2} \nonumber\\
&\times\left[ (\lambda-1){\rm cos}(k\eta-\theta) + \sqrt{-D} {\rm sin}(k\eta-\theta) \right],
\end{align}
where $\theta$ is defined by $\theta\equiv \frac{1}{2}\sqrt{-D}{\rm log}(\tau/\tau_0)$. 
The phase velocity of the fluctuation is $\sqrt{-D}/(2k\tau)$.

We set the initial time to $\tau_0=1$ fm and use the ideal gas equation of state, $e=3p$ and $\lambda=1/3$. 
The initial energy density $e_0$ in Bjorken flow is given by  $e_0= 1000$ ${\rm fm}^{-4}$. 
In the case of $D>0$, we choose $k=0.5$, $D=0.111$ and $A=0.1$ ${\rm fm}^{-4}$. 
In the case of $D<0$, we choose $k =2\pi$, $D=-52.193$ and $A=0.1$ ${\rm fm}^{-4}$.

Figures \ref{fig:sounddamp}  and \ref{fig:sound} show the analytical and numerical results of the fluctuations of the energy density and the velocity around Bjorken flow for $D>0$  and $D<0$. 
In the numerical calculation, we use the periodic boundary condition. 
When the value of $D$ is positive, the fluctuation does not propagate and its amplitude decreases with time (Fig.\:\ref{fig:sounddamp}). 
On the other hand, when the value of $D$ is negative, the fluctuation propagates and its amplitude decreases with time (Fig.\:\ref{fig:sound}). 
However, if we chose different initial conditions such that more than one mode were involved, the amplitude of the fluctuation would at first grow and then reduce for both $D>0$ and $D<0$ due to the interference of two modes. 
The similar amplification of the fluctuations around  Bjorken flow is reported in Ref.\:\cite{Floerchinger2011}. 
Our numerical results show good agreement with the analytical solutions.

\subsection{Gubser flow} \label{sec:Gubser}

\begin{figure}[t!]
  \centering
\includegraphics[width=8cm]{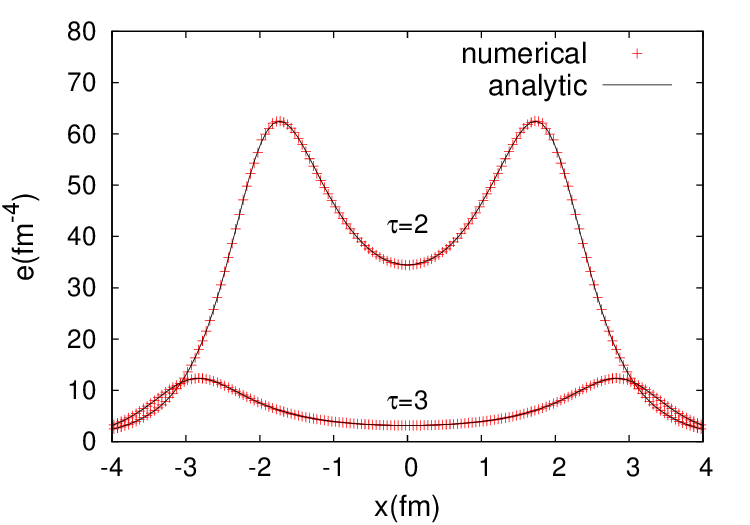}
  \centering
  \includegraphics[width=8cm]{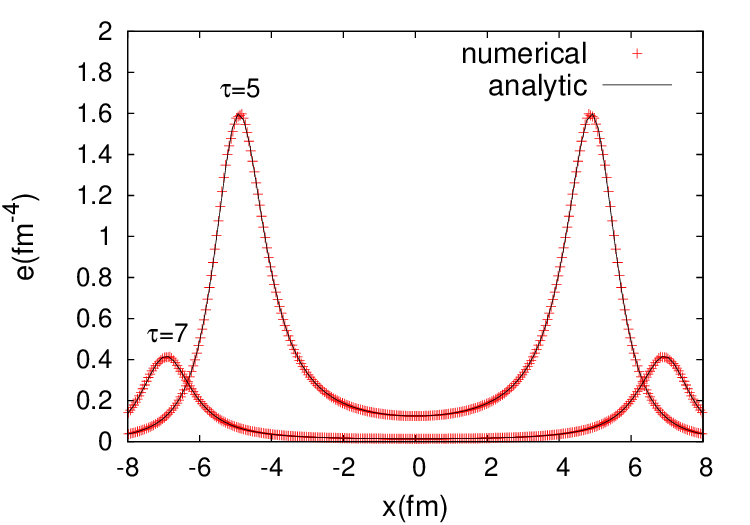}
  \caption{Comparison between the analytical solution and the numerical calculation of the energy density $e$ for the Gubser flow. Top: At  $\tau=2$ and 3 fm, Bottom: At $\tau=5$ and 7 fm. \label{fig:Gubser_e}}
\end{figure}
\begin{figure}[t!]
  \centering
  \includegraphics[width=8cm]{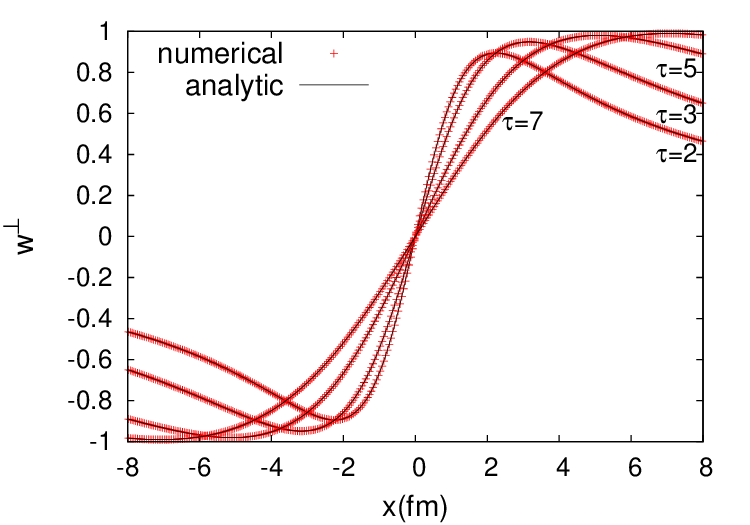}
  \caption{Comparison between the analytical solution and the numerical calculations of transverse velocity $w^\bot$ for the Gubser flow at $\tau=2,3,5$ and 7 fm.
  }
  \label{fig:Gubser_v}
 \end{figure}

An analytic solution to the relativistic, conformally invariant Navier-Stokes equation is constructed based on symmetry considerations in $(\tau, \eta, x_\bot, \phi)$ coordinate system \cite{Gubser2010,Gubser2011}.  
The Gubser flow and its related solutions are utilized for checking or improvement of hydrodynamic codes \cite{Pang:2014ipa, Denicol2013, ECHO2013, Karpenko2014, Noronha2015, Shen2014, ECHO2015}. 
The solution is a generalization of Bjorken flow where the medium expands both longitudinally and radially, which gives us a realistic description of the space-time evolution of high-energy heavy-ion collisions. 
According to the solution for inviscid fluid, the transverse or radial velocity $w^\bot$ and energy density $e$ are given by 
\begin{equation} w^\bot =\frac{u^\bot}{u^\tau}=\frac{2q^2\tau x_\bot}{1+q^2\tau^2+q^2x_\bot^2} , \label{eq:gubserv} 
\end{equation}
\begin{equation} e =\frac{\hat{e_0}}{\tau^{4/3}}\frac{(2q)^{8/3}}{[1+2q^2(\tau^2+x_\bot^2)+q^4(\tau^2 -x_\bot^2 )^2]^{4/3}}, \label{eq:gubsere}
\end{equation}
where $\hat{e_0}$ is a dimensionless integration constant, $q$ is an arbitrary dimensional constant with unit of inverse length of the system size \cite{Gubser2010,Gubser2011}.

\par We compare our numerical calculations and the analytical solution in Figs.\ref{fig:Gubser_e}  and \ref{fig:Gubser_v}.
In our numerical calculation,  the parameters are set to  $q=1$ fm$^{-1}$ and $\hat{e_0} =400$. 
The hydrodynamic expansion starts at $\tau_0=1$ fm.
The numerical simulation is performed with the grid size $\Delta x=\Delta y=0.05$ fm, $\Delta\eta=0.1$ and the time-step size $\Delta \tau= 0.1 \Delta x$. 
First the consistency between our numerical calculations and the analytical solution suggests the Strange splitting method in Sect.\:\ref{sec:3d} works correctly.
In the Gubser flow, the existence of the initial transverse flow is an origin of the strong transverse flow at later time.
To reproduce the strong transverse flow in the Gubser flow, 
we find that careful choice of  the interpolation method is required in the numerical calculation.
For example, if the second-order interpolation method is employed, the energy density around $x=0$ fm of the numerical calculations 
is larger than that of the analytical solutions.
However, if we use the PPM interpolation method, we can reproduce the analytical solutions numerically.
It is discussed in \ref{app-Gubser}.

\begin{table*}[hbtp]
  \centering
  \begin{tabular}{cc|cc|cc}
    \hline
    &   &  \multicolumn{2}{|c|}{conservative form} & \multicolumn{2}{|c}{with source term} \\
    $\Delta\eta$ & $E_0\;({\rm GeV})$  & $\varepsilon _E$ & $\sum|M^{n+1}-M^n| \;({\rm GeV})$ &  $\varepsilon_E$  &  $\sum|M^{n+1}-M^n|\; ({\rm GeV})$ \\
    \hline 
    0.02 & 1117 &   7.48E-10   &  6.00E-07 &  6.42E-04 &  2.02E-02 \\
    0.1   & 1117 &    2.85E-10  & 1.68E-07 &  3.77E-03  & 1.10E-01 \\
    0.2   &1133 &   6.46E-10 & 1.65E-07    & 7.48E-03 &  7.90E-02 \\
    0.5   & 1148 &  1.48E-10 & 7.17E-08   &  4.36E-02   &1.60E-01 \\
    \hline
  \end{tabular}
   \caption{The violation of the total energy and momentum conservation.}
   \label{table:conservation}
\end{table*}

\section{Conservation property}
We check the energy and momentum conservation of our code. 
The conserved quantities in our algorithm in Milne coordinates are given by $\tau T^{\tau \nu} (\nu = t,x,y,z)$. 
Their time evolution in our algorithm is schematically written by 
\begin{equation} 
\tau^{n+1}(T^{\tau \nu})^{n+1}_i = \tau^n(T^{\tau \nu})^{n}_i - \frac{\Delta\tau}{\Delta\eta}(F^\nu_i - F^\nu_{i-1}), 
\label{eq:ec1}
\end{equation}
where $F^\nu_i$ represents the flux of the conserved quantities which flow into and out of $i$-th grid cell during $\Delta\tau$. 
Equation \eqref{eq:ec1} contains two steps: Lagrange and remap steps.
Integrating Eq.\:\eqref{eq:ec1} on all spatial grids, we obtain  
\begin{align}
 \sum_{i=i_{min}}^{i_{max}} \tau^{n+1} (T^{\tau \nu})^{n+1}_i =& \sum_{i=i_{min}}^{i_{max}}\tau^n(T^{\tau \nu})^{n}_i  \nonumber\\
&- \frac{\Delta\tau}{\Delta\eta}(F^\nu_{i_{max}} - F^\nu_{i_{min}-1}),  \label{eq:hozon}
\end{align}
which suggests that the total variation of the conserved quantities depends on the amount of inflow and outflow from the boundary. 
If the equations with the source terms, Eqs.\:\eqref{eq:1}--\eqref{eq:4} are used in the code, the right-hand side of Eq.\:\eqref{eq:hozon} has an additional term from the source terms, which can spoil the conservation property and affects the numerical accuracy in application to physical problems.

\par Here we focus on the effects of existence of the source terms in Milne coordinates on conservation property. 
We perform our numerical calculation with the initial energy density and flow distributions which are usually used in study of the relativistic heavy-ion collisions, 
\begin{align}
e(\tau_0, \eta ) =& e_0{\rm exp}\left[-\frac{(|\eta_s| - \eta_{\rm flat}/2)^2}{\sigma_\eta^2} \theta( |\eta| - \eta_{\rm flat}/2) \right] \nonumber \\
& \times\theta(Y_b - |\eta|) \label{eq:flat_e},\\
w^\eta(\tau_0, \eta) =& 0 \label{eq:flat_v}, 
\end{align}
where $Y_b=5.3$ is the beam rapidity, $\sigma_\eta=2.1$ and $\eta_{\rm flat}=2.6$ show the size of the flat structure of the initial energy density distribution in the rapidity, and $e_0=30\;{\rm GeV/fm}^3$ is the maximum value of the energy density. 
We choose a typical parameter set which is tuned for the RHIC collision energy \cite{Hirano2006, Karpenko2014}.  
To discuss the source-term effect in Milne coordinates clearly,  we carry out the numerical calculation only in the rapidity direction. 
In Ref. \cite{Karpenko2014} one calculates the total energy and entropy in the beginning and at the end of 3D hydrodynamic evolution, using the Glauber model with a limited rapidity profile for an initial condition.
From comparison between them they find that the energy is conserved on a level of better than 3 $\%$ in their code. 

\par To check the conservation property of our code, we evaluate the total energy and momentum in the hydrodynamic expansion, 
\begin{align}
E(\tau^n ) & = \tau^n \Delta\eta \sum_{\rm all \;grid} (T^{\tau 0}) ^n_i,  \\
M(\tau^n) & =  \tau^n \Delta\eta \sum_{\rm all \; grid} (T^{\rm \tau z})^n_i. 
\end{align}
We carry out a numerical calculation from  $\tau_0=1$ fm to $\tau = 10$ fm on $\Delta\eta=0.02, 0.1, 0.2,$ and 0.5 grid sizes with $\Delta\tau = 0.1 \tau_0 \Delta\eta$ time-step size. 
The total momentum in the beginning is vanishing in the numerical precision. 
We observe that in the case where the source terms are explicitly included, the total energy and momentum increase with the proper time $\tau$ monotonically on $\Delta \eta=0.5$. 
On other grid sizes, first the total energy and momentum increase with $\tau$ and after some time steps they start to decrease. 
If there are no source terms, sometime the total energy and momentum increase and at other times they decrease. 
This behavior suggests that the simple comparison between total conserved quantities in the beginning and those at the end is not suitable for an investigation of the conservation property. 
Instead of the simple comparison, we evaluate the violation of the energy and momentum conservation at each time step and sum it from the beginning to the end, 
\begin{align} 
\varepsilon _E & \equiv \sum_{\rm all\; step} \frac{|E(\tau^n) - E(\tau^{n-1})|}{E(\tau_0)},  \\
\varepsilon _M & \equiv \sum_{\rm all\; step} \frac{|M(\tau^n) - M(\tau^{n-1})|}{|M(\tau_0)|}. 
\end{align}
We show the calculated results of them in Table\: \ref{table:conservation}. 
We find that the numerical calculation based on the equations without the source terms keeps the energy and momentum conservation with high accuracy compared with that with source terms on every grid size. 
In the case where the source terms are explicitly included, 
enough numerical accuracy is still kept, but the amount of the violation of the conservation property increases with grid size. 
On the other hand, in the numerical algorithm with the conservative form, it does not depend on the grid size. 
Even on the the course grid, the conservation property is kept with very high accuracy.

\begin{figure}[t!]
  \centering
  \includegraphics[width=8cm]{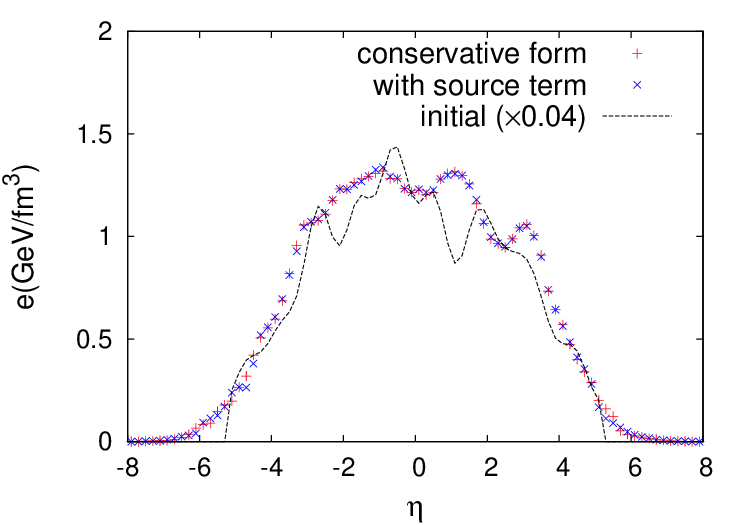}
  \centering
  \includegraphics[width=8cm]{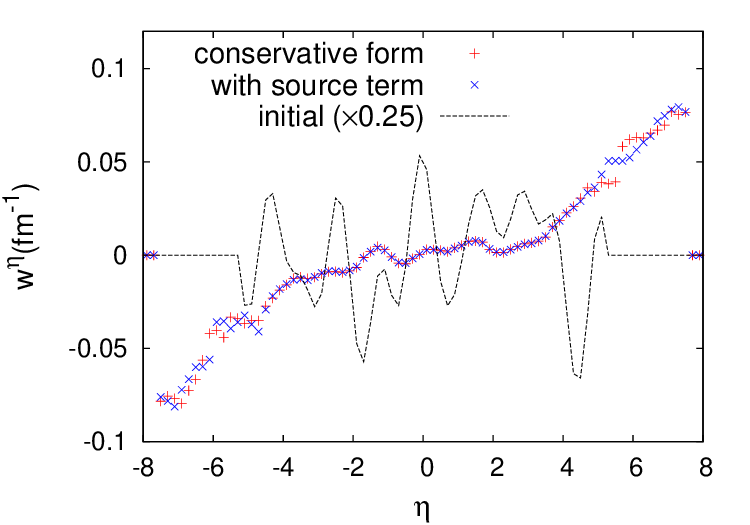}
  \caption{The numerical results for longitudinal expansion with the fluctuating initial condition at $\tau =10$ fm. 
  Top panel: The energy distributions from the codes with the conservative form and with the source terms. 
  The dotted line stands for the initial energy distribution.
  Bottom panel: The velocity distribution from the codes with the conservative form and with the source terms. 
 The dotted line stands for the initial velocity distribution. \label{fig:conservation}}
\end{figure}

\par Next we investigate the total energy and momentum conservation in hydrodynamic evolution which starts from a fluctuating initial condition. 
We add the fluctuation to the energy density and velocity distributions in Eqs.\:\eqref{eq:flat_e} and  \eqref{eq:flat_v}, 
\begin{align}
e(\tau_0,\eta) =& e^{\rm flat}(\tau_0,\eta)  \nonumber \\ 
&\times\left( 1+\sum_{n=0}^{10} \delta e_n {\rm cos} \left(n\frac{2\pi(\eta - \eta_n^e)}{L}\right)\right),  \\
w^\eta(\tau_0,\eta) =& \sum_{n=0}^{10}\delta w^\eta_n {\rm cos}\left(n\frac{2\pi(\eta - \eta_n^v)}{L}\right), 
\end{align}
where $e^{\rm flat}$ is given by Eq.\:\eqref{eq:flat_e} and values of $\eta^e_n$ and $\eta^v_n$ are chosen between $\eta=-Y_b$ and $Y_b$ at random. We set $\delta e_n=0.05$ and  $\delta w^\eta=0.05$ fm$^{-1}$ for all $n$. 
We carry out numerical calculations with the grid size $\Delta\eta=0.2$, which is often chosen in calculations of high-energy heavy-ion collisions \cite{Karpenko2014}.
We set the time-step size equal to $\Delta\tau =0.1\tau_0\Delta\eta$.
In Fig.\:\ref{fig:conservation} the energy density and velocity distributions at $\tau=10$ fm with and without the source terms are shown. 
At the mid-rapidity the energy density and the flow distributions of numerical calculations with and without the source terms are consistent with each other. 
In the region of $|\eta|>4$, however,  the differences between them are observed in the small structure of both distributions. 
The growth of the velocity to the vacuum $|\eta| \sim 8$ gives a difficulty of numerical calculation and becomes the reasons for the differences. 

The deviation from the energy and momentum conservation are listed in Table\:\ref{table:fluctuation}. 
For both cases, we find  that they are around ten times as large as those with the smoothed initial condition. 
Nevertheless, the code based on the conservative form keeps conservation property with high accuracy. 
On the other hand, in the code with the source terms a few \% deviation from the energy and momentum conservation appears, which is still acceptable. 
In the code with the source terms, numerical calculation with fine grid size is indispensable for the energy and momentum conservation. 
There exist other ingredients which can cause additional error, and violation of the conservation property originates from the geometric source term, for instance shock waves and jets 
in medium \cite{Chaudhuri:2005vc, Betz:2008js,Tachibana:2014lja}.
In addition, the existence of the viscosity can be the origin of the breakdown of the conservation property \cite{Karpenko2014}.
To avoid such problems, we need to construct the codes based on constitutive equations with the conservative form or perform numerical calculations on 
sufficiently fine grids. 

\begin{table}[t!]
  \centering
  \begin{tabular}{c|cc}
    \hline
    &  $\varepsilon_E$ & $ \varepsilon _M$   \\
    \hline 
     conservative & 1.38E-09 &   8.59E-09 \\
     with souce & 1.27E-02 &   5.61E-02 \\
    \hline
  \end{tabular}
  \caption{The violation of the total energy and momentum conservation with fluctuating initial conditions. The initial total energy and momentum are $E_0=2224$ GeV and $M_0=-94$ GeV.\label{table:fluctuation}}
\end{table}

\section{Summary}
We constructed a new Godunov type relativistic hydrodynamic code in Milne coordinates based on the algorithm in Cartesian coordinates \cite{Akamatsu2014}. 
We evaluated the flux terms, using the numerical solution of the  Riemann problem with the initial condition at the constant proper time $\tau$. 
We checked the correctness of our algorithm from the comparison between numerical calculations and analytical solutions of  shock tube, 
expansion of matter into the vacuum, the Landau-Khalatnikov solution, propagation of fluctuation around Bjorken flow and the Gubser flow. 
We investigated the energy and momentum conservation of our code from a calculation of the longitudinal hydrodynamic expansion with an 
initial condition for high-energy heavy-ion collisions. 

In particular, we focused on the effects of the source terms in relativistic numerical hydrodynamics in Milne coordinates on stability and numerical viscosity.
We analyzed those effects in the test problems of expansion into the vacuum and the conservation property.  
In expansion of matter into the vacuum, we showed that numerical results from the code without the source terms are closer to the analytical solution compared with that with source terms. 
Besides, the code without the source terms is more stable  and  has less  numerical viscosity than the code with the source terms.  
In addition, we observed that the code written in the conservative form keeps the conservation property with high accuracy 
in the expansion from the fluctuating initial longitudinal profile for high-energy heavy-ion collisions, even on a coarse grid.

Our algorithm is easily extended to the code with the QCD equation of state and finite viscosities \cite{Akamatsu2014, Inutsuka2011}.
After that, we shall employ our hydrodynamic code to investigate experimental results at RHIC and LHC and understand the detailed  QGP bulk property using a reliable 3D relativistic viscous hydrodynamic expansion with small numerical viscosity. 

\section*{Acknowledgments}
The work of CN is supported by the JSPS Grant-in-Aid for Scientific Research (S) No. 26220707 and 
US Department of Energy Grant DE-FG02-05ER41367. 
Y. A. is partially supported by JSPS Postdoctoral Fellowships for Research Abroad.

\appendix

\def\thesection{Appendix \Alph{section}} 

\section{Riemann problem in Milne coordinates} \label{app-proof}
We show that the two initial problems Eqs.\:\eqref{eq:riemann} and \eqref{eq:initial} have the same analytic solution.  
First, we prove that the hydrodynamic state which satisfies the condition $\partial_\eta \bm V=0$ at some proper time keeps $\partial_\tau \bm V=0$. 
In other words, if the hydrodynamic state $\bm V$ is uniform at some initial proper time, it remains the initial uniform state at all times.
This state corresponds to the initial condition of the Riemann problem in Milne coordinates Eq.\:\eqref{eq:initial}. 
For the $(1+1)$-dimensional case, we rewrite the energy conservations, Eqs.\:\eqref{eq:7} and \eqref{eq:8}, as 
\begin{align} &\partial_\tau (\tau {\rm cosh\eta}\:T^{tt} -\tau {\rm sinh\eta}\: T^{tz} ) \nonumber\\
 &\quad + \partial_\eta (-{\rm sinh}\eta\: T^{tt}+{\rm cosh}\eta\: T^{tz}) =0, \label{eq:steady1}\\
& \partial_\tau(\tau{\rm cosh}\eta\: T^{tz} -\tau {\rm sinh}\eta\: T^{zz}) \nonumber \\
& \quad +\partial_\eta (-{\rm sinh}\eta \:T^{tz}+{\rm cosh}\eta \:T^{zz}) =0 . \label{eq:steady2} 
\end{align}
Inserting the conditions, $\partial_\eta \bm V=0$, namely $\partial_\eta T^{tt} =0$, $\partial_\eta T^{tz}=0$ and $\partial_\eta T^{zz}= 0$, into Eqs.\:\eqref{eq:steady1}, \eqref{eq:steady2}, and the derivative of Eq.\:\eqref{eq:steady1} or \eqref{eq:steady2} with respect to rapidity, we obtain 
\begin{equation}\partial_\tau T^{tt}=0,\quad \partial_\tau T^{zz}=0, \quad \partial_\tau T^{tz}=0, 
\end{equation}
which means $\partial_\tau\bm V=0$.  
This result indicates that in the Riemann problem in Milne coordinates Eq.\:\eqref{eq:initial}, the hydrodynamic state 
outside the light cone of the discontinuity $(\tau_0, \eta_i)$ remains $\bm V_L$ or $\bm V_R$, based on the fact that the signal velocity in the ideal hydrodynamic equation is smaller than the speed of light. 

\begin{figure}[t]
  \centering
  \includegraphics[width=7cm]{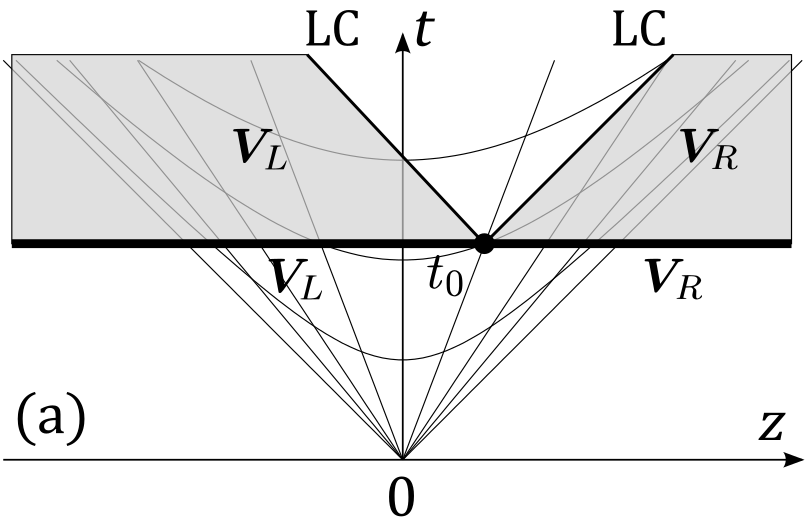}
  \centering
  \includegraphics[width=7cm]{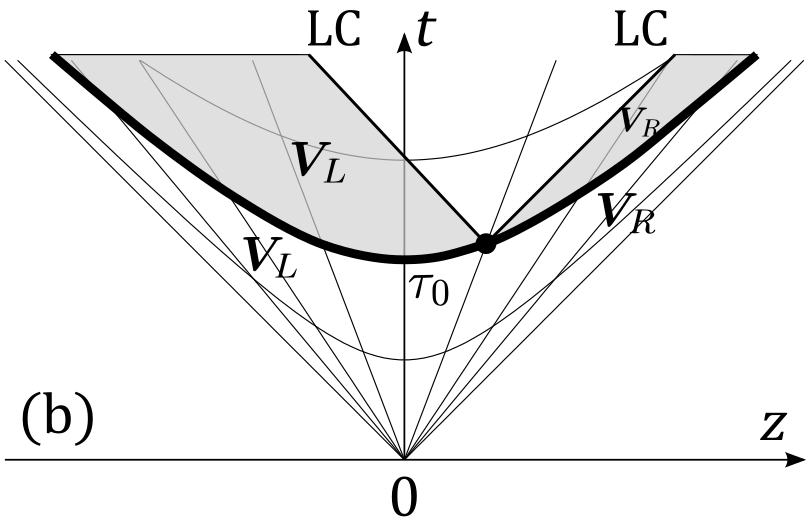}
  \caption{
  The Riemann problem in Cartesian coordinates (a) and that in Milne coordinates (b). 
The solid circle stands for the initial discontinuity. 
The thick solid lines indicate initial conditions of the Riemann problem in 
Cartesian coordinates (a) and Milne coordinates (b).
LC stands for the light cone of the discontinuity. } 
  \label{fig:finite-etai}
\end{figure}

In Fig.\:\ref{fig:finite-etai}\:(b), the initial condition of the Riemann problem is set on the hyperbolic curve $\tau=\tau_0$ with the discontinuity at $\eta_i$, the hydrodynamic state outside the light cone of the discontinuity is given by $\bm V_L$ or $\bm V_R$.
Figure \ref{fig:finite-etai}\:(a) shows the initial condition of the Riemann problem in Cartesian coordinates with the same discontinuity point at $(t_0,z_i)$, which corresponds to $(\tau_0, \eta_i)$ in Fig.\:\ref{fig:finite-etai}\:(b).
From a comparison between Figs.\:\ref{fig:finite-etai}\:(a) and (b), the analytical solutions for initial-condition problems which are given by Eqs.\:\eqref{eq:riemann} and \eqref{eq:initial} 
satisfy the same boundary conditions on the light cone at the discontinuity $(\tau_0,\eta_i)$. 
We conclude that the analytical solution of the Riemann problem in Milne coordinates, Eq.\:\eqref{eq:initial} 
is the same as that in Cartesian coordinates, Eq.\:\eqref{eq:riemann}.

\section{Discretized hydrodynamic equation with Lagrange step} \label{app-flux}

We show  the detailed calculation of numerical flux terms Eqs.\:\eqref{eq:flux1}-\eqref{eq:flux3}.
We represent the velocity and pressure of the grid-cell boundary at $\eta_i$ as $V^z_i$ and $P_i$, respectively. 
First we shift the grid-cell boundary at $\eta_i$ to $\eta=0$ by the Lorentz boost transformation to derive the moving distance $\delta\eta_i(\tau')$ of the grid-cell boundary during $\tau'(0\leq \tau' \leq \Delta\tau)$. 
Under the Lorentz boost transformation, $\delta\eta_i(\tau')$ is invariant. 
The velocity of the grid-cell boundary $V^z_{0,i}$  after the Lorentz boost transformation is related to $V^z_i$ 
by 
\begin{equation}
V^z_i=\frac{{\rm sinh}\eta_i+ V^z_{0,i}{\rm cosh}\eta_i}{{\rm cosh}\eta_i+V^z_{0,i}{\rm sinh}\eta_i}.
\end{equation}
The moving distance $\delta\eta_i(\tau')$ of the grid-cell boundary from $\eta=0$ during $\tau'$ is given by Eq.\:\eqref{eq:deltaeta} with Eq.\:\eqref{eq:deltat}.
Equation \eqref{eq:deltat} is derived from the simultaneous equations $z=V^z_{0,i} (t - \tau^n)$ and $(\tau^n+\tau')^2=t^2 - z^2$ with $t=\tau^n+\delta t(\tau')$. 

Next we 
evaluate the numerical flux term of Eq.\:\eqref{eq:evolve} on the original frame (where the grid-cell boundary moves from $\eta_i$).
In the integration of Eq.\:\eqref{eq:evolve}, the unit vector $n_{\alpha, i}$ is written by 
\begin{equation}
 n_{\alpha,i}=\gamma_i ({\rm sinh}\eta -V_i^z{\rm cosh}\eta,\; \tau({\rm cosh}\eta -V_i^z{\rm sinh}\eta ) )
\end{equation}
with $\gamma_i=\{1- (V_i^z)^2\}^{-1/2}$. The integrands of the flux terms become
\begin{align} &T^{\alpha t}n_{\alpha,i} = \gamma_i P_i V_i^z, \\
& T^{\alpha z} n_{\alpha, i} = \gamma_i P_i , \\
& T^{\alpha j} n_{\alpha,i}= 0, \qquad(j =x, y).
\end{align}
The integral element $ds$ on $C_i$ is given by 
\begin{equation} 
ds =\gamma^{-1}_i \frac{d\tau'}{{\rm cosh}(\eta_i+\delta\eta_i(\tau')) -V^z_i{\rm sinh}(\eta_i +\delta\eta_i(\tau') )}.  \label{eq:ds}
\end{equation}
Expanding Eq.\:\eqref{eq:ds} in terms of $\tau'$ with Eq.\:\eqref{eq:deltaeta}, we perform the integration of $\tau'$ in $0\leq \tau' \leq \Delta\tau$. Then we can get the explicit form of the numerical flux terms as Eqs.\:\eqref{eq:flux1}-\eqref{eq:flux3}.

\section{Riemann problem from Cartesian to Milne coordinates} 
\label{app-Riemann-taueta}
 We represent the analytical solution of the Riemann problem in Cartesian coordinates \cite{Marti1994, Pons2000}, 
as that in Milne coordinates. 
The analytical solution is composed of the four uniform hydrodynamic states, $\bm V_L, \bm V_{L^*}, \bm V_{R^*}$,
and $\bm V_{R}$, which are separated by 
three nonlinear waves; two of them are shock waves and/or rarefaction waves, the other is a contact discontinuity which separates $\bm V_{L^*}$ and $\bm V_{R^*}$. 
The four hydrodynamic states and the three nonlinear waves move with constant velocity.
Suppose that the initial discontinuity in Milne coordinates is located at $(\tau_0,\eta_i)$. 
At proper time $\tau_f=\tau_0+\Delta\tau$, the location of the discontinuity surface of the shock front, rarefaction front, and contact discontinuity, which move with constant velocity $V^z$ from $\eta_i$, $\eta_f$ is written
\begin{equation} \eta_{f} = \frac{1}{2}{\rm log}\frac{ (1+V^z) \delta t + \tau_0 {\rm cosh}\eta_i + \tau_0 {\rm sinh}\eta_i}{(1-V^z) \delta t + \tau_0 {\rm cosh}\eta_i - \tau_0{\rm sinh}\eta_i },  \label{eq:etafinal}
\end{equation}
where $\delta t$ is given by 
\begin{align} &\delta t = \frac{1}{1- (V^z)^2} \Biggl\{ - \tau_0({\rm cosh}\eta_i - V^z{\rm sinh}\eta_i ) \nonumber\\
&\qquad +\Bigl( \tau_0^2({\rm cosh}\eta_i - V^z {\rm sinh}\eta_i)^2 \nonumber\\
&\qquad\quad+ (1 - (V^z)^2)(\Delta\tau^2 + 2\tau_0 \Delta\tau) \Bigr)^{1/2} \Biggr\}.
\end{align}

\par The state of the rarefaction wave depends on $\xi=(z-z_i)/ (t-t_0)$ in Cartesian coordinates, where $t_0$ is the initial time and $z_i$ is 
the location of the initial discontinuity. 
Here $t_0$ and $z_i$ are related with $\tau_0$ and $\eta_i$  in Milne 
coordinates through the relations 
$\tau_0=(t_0^2 - z_i^2)^{1/2}$ 
and $\eta_i ={\rm tanh}^{-1} (z_i/t_0)$. 
In Milne coordinates, the state of the rarefaction wave is described by 
\begin{equation} \xi = \frac{ \tau{\rm sinh}\eta -\tau_0{\rm sinh}\eta_i}{\tau{\rm cosh}\eta -\tau_0{\rm cosh}\eta_i}. 
\end{equation}

In the case of the expansion of matter into the vacuum, the location of the boundary to the matter at rest in Cartesian coordinates is obtained by 
Eq.\:\eqref{eq:etafinal}, where $V^z$ is replaced by negative sound velocity, $-c_s$. 
The boundary to the vacuum which moves with the speed of light is given by 
\begin{equation} \eta_f = \eta_i+\frac{1}{2}{\rm log}\left(\frac{\tau_f}{\tau_0}\right). 
\end{equation}

\section{General solution of longitudinal fluctuations around Bjorken flow}\label{app-general-solution}
We derive analytical solutions for fluctuations around Bjorken flow \cite{Bjorken1983}.
We start from the $(1+1)$-dimensional relativistic ideal hydrodynamic equations in Milne coordinates, 
\begin{align} &(\partial_\tau + w^\eta\partial_\eta) (\tau e) = - \frac{\tau(e+p)}{u^\tau} \partial_\mu u^\mu \nonumber\\
&\qquad\qquad\qquad\qquad + \tau^2 (e+p) (u^\eta)^2 - p, \label{eq:Euler1}\\
& (\partial_\tau + w^\eta\partial_\eta ) w^\eta = -\frac{1}{(u^\tau)^2 (e+p)} \left(\frac{1}{\tau^2}\partial_\eta p + w^\eta \partial_\tau p\right) \nonumber\\
&\qquad\qquad\qquad\qquad + \tau (w^\eta)^3 - \frac{2}{\tau} w^\eta, \label{eq:Euler2}
\end{align}
which are obtained from Eqs.\:\eqref{eq:tensor}, \eqref{eq:source3}, and \eqref{eq:source4}.
Focusing on the propagation of fluctuations around Bjorken flow, we put a small perturbation on top of the Bjorken expansion in 
Eqs.\:\eqref{eq:Euler1} and \eqref{eq:Euler2}, 
\begin{equation} e=e_B+\delta e,\quad w^\eta =\delta w^\eta,\quad e_B=e_0\left(\frac{\tau_0}{\tau}\right)^{1+\lambda}, 
\end{equation}
where $e_B$ is the energy density from Bjorken's scaling solution  \cite{Bjorken1983},  
the equation of state is given by $p=\lambda e$, the sound velocity is $c_s = \sqrt{\lambda}$. 
Assuming that the fluctuations are small, we neglect the second and higher order terms of  fluctuations in Eqs.\:\eqref{eq:Euler1} and \eqref{eq:Euler2} 
and derive the linearized relativistic hydrodynamic equations, 
\begin{align} &\partial_\tau \delta e+ (1+\lambda)e_B \partial_\eta \delta w^\eta+\frac{1+\lambda}{\tau}\delta e =0,  \\
& \partial_\tau \delta w^\eta +\frac{\lambda}{1+\lambda}  \frac{1}{\tau^2 e_B}\partial_\eta \delta e+\frac{2-\lambda}{\tau} \delta w^\eta =0. 
\label{AppEq:Euler-fluc}
\end{align}

We input the following initial conditions for  $\delta e$ and $\delta w^\eta$ at $\tau=\tau_0$, 
\begin{align} \delta e(\tau_0,\eta) &= A_1 e^{ik\eta},\label{eq:sound-initial1}\\
 \delta w^\eta(\tau_0, \eta) &= A_2 e^{ik\eta}.  \label{eq:sound-initial2}
\end{align}
Using the Fourier transform of $ \delta e (\tau, \eta)$ and $\delta w^\eta (\tau,\eta)$,  
\begin{align}& \delta e (\tau, \eta) = \int\frac{dk}{2\pi} \delta\tilde{e}(\tau, k) e^{ik\eta} , \\ 
&\delta w^\eta (\tau,\eta)= \int\frac{dk}{2\pi} \delta \tilde{v}^\eta(\tau,k)e^{ik\eta}, 
\end{align}
we obtain the analytical solutions, 
\begin{align} \delta e(\tau,\eta)&=  \frac{1}{\sqrt{D}} \left(\frac{\tau}{\tau_0}\right)^{-(3+\lambda)/2} \nonumber\\
& \times \Biggl[\left( a_1A_1 +  i k e_0(1+\lambda) \tau_0A_2 \right)\left(\frac{\tau}{\tau_0}\right)^{-\sqrt{D}/2} \nonumber\\ 
& -  \left(a_2A_1  + i k e_0(1+\lambda)\tau_0A_2 \right) \left(\frac{\tau}{\tau_0}\right)^{\sqrt{D}/2} \Biggr] e^{ik\eta} \label{eq:deltae}, \\
\delta w^\eta(\tau,\eta) &= \frac{1}{\sqrt{D}}\left(\frac{\tau}{\tau_0}\right)^{(\lambda-3)/2}\nonumber\\
&\times \Biggl[  \left(- a_2A_2 + i\frac{k\lambda A_1}{\tau_0e_0(1+\lambda)} \right) \left(\frac{\tau}{\tau_0}\right)^{-\sqrt{D}/2} \nonumber \\
&+ \left( a_1A_2 - i\frac{k\lambda A_1}{\tau_0e_0(1+\lambda)} \right) \left(\frac{\tau}{\tau_0}\right)^{\sqrt{D}/2}\Biggr] e^{ik\eta} \label{eq:deltav}, 
\end{align}
where $a_1$, $a_2$ and $D(\neq 0)$ are given by 
\begin{equation} a_1= \frac{1}{2}\left( \lambda-1 + \sqrt{D}\right),\quad a_2=\frac{1}{2}\left(\lambda-1 - \sqrt{D}\right),  \label{eq:a1a2}
\end{equation}
\begin{equation}D = (1-\lambda)^2-4k^2\lambda, \label{eq:sound-D}
\end{equation}
respectively. 
Equation \eqref{eq:sound-D} indicates that the value of $D$ is always less than $(1-\lambda)^2$ and decreases with the wave number $k$. 
As a result, the first and second terms of Eqs.\:\eqref{eq:deltae} and \eqref{eq:deltav} have a negative power of $\tau$, which 
indicates that the amplitude of the fluctuation decreases with $\tau$. 

In the case of $D>0$, the initial fluctuations are attenuated without propagation. 
In the particular cases of Eq.\:\eqref{eq:deltae} and \eqref{eq:deltav}, if  $A_1$ and $A_2$ satisfy the 
following relation: 
\begin{equation}  a_2A_1+ik e_0(1+\lambda)\tau_0A_2 = 0 \label{eq:constraint}, 
\end{equation}
then only the first-term mode in Eqs.\:\eqref{eq:deltae} and \eqref{eq:deltav} remains. 
The analytical solutions become  
\begin{align} \delta e(\tau,\eta)&= A_1\left(\frac{\tau}{\tau_0}\right)^{(-3-\lambda-\sqrt{D})/2} e^{ik\eta} \label{eq:soundedamp}, \\
\delta w^\eta(\tau,\eta) &= A_2\left(\frac{\tau}{\tau_0}\right)^{(-3+\lambda-\sqrt{D})/2}e^{ik\eta},  \label{eq:soundvdamp}
\end{align}
where the fluctuations with the smaller wave number attenuate faster. 

In the case of $D<0$, 
the first and the second terms of Eqs.\:\eqref{eq:deltae} and \eqref{eq:deltav} represent the progressive 
and regressive waves, respectively.  
If  the coefficients $A_1$ and  $A_2$ satisfy Eq.\:\eqref{eq:constraint}, 
then the analytical solutions become
\begin{align} \delta e(\tau,\eta) &= A_1\left(\frac{\tau}{\tau_0}\right)^{-(3+\lambda)/2} e^{i(k\eta -\theta)} \label{eq:soundeprop}, \\
\delta w^\eta (\tau,\eta) &= A_2\left(\frac{\tau}{\tau_0}\right)^{(\lambda-3)/2}e^{i(k\eta-\theta)},  \label{eq:soundvprop}
\end{align}
where $\theta$ is defined by $\theta\equiv \frac{1}{2}\sqrt{-D}{\rm log}(\tau/\tau_0)$. 
The propagating speed of the fluctuations is $\sqrt{-D}/(2k\tau)$.
The condition of $D<0$ is given by 
\begin{equation}  k > \frac{1- c_s^2}{2 c_s}, 
\end{equation}
where the fluctuation with the larger wave number propagates faster. 
In the case of  $D=0$, the analytical  solutions are given by 
\begin{align} &\delta e(\tau,\eta) =  \left(\frac{\tau}{\tau_0}\right)^{-(3+\lambda)/2}\nonumber \\
&\times \left[ A_1+\left( \frac{1-\lambda}{2}A_1 - ik e_0(1+\lambda)\tau_0A_2 \right){\rm log}\frac{\tau}{\tau_0} \right] e^{ik\eta} \label{eq_e_d0},  \\
&\delta w^\eta(\tau,\eta)  =   \left(\frac{\tau}{\tau_0}\right)^{(\lambda-3)/2} \nonumber \\
&\times \left[ A_2 - \left( \frac{1-\lambda}{2}A_2 + i \frac{k\lambda A_1}{\tau_0 e_0 (1+\lambda)}\right){\rm log}\frac{\tau}{\tau_0} \right] e^{ik\eta}  \label{eq_v_d0}. 
\end{align}
If  $A_1$ and $A_2$ satisfy 
\begin{equation} \frac{1-\lambda}{2}A_1 = ik e_0(1+\lambda)\tau_0A_2, 
\end{equation}
then the analytical solutions, Eqs.\:\eqref{eq_e_d0} and \eqref{eq_v_d0}, become 
\begin{align} \delta e(\tau,\eta) &= A_1 \left(\frac{\tau}{\tau_0}\right)^{-(3+\lambda)/2} e^{ik\eta}, \\
\delta w^\eta (\tau,\eta) &= A_2\left(\frac{\tau}{\tau_0}\right)^{(\lambda-3)/2}e^{ik\eta}. 
\end{align}

\begin{figure}[b]
  \centering
  \includegraphics[width=8cm]{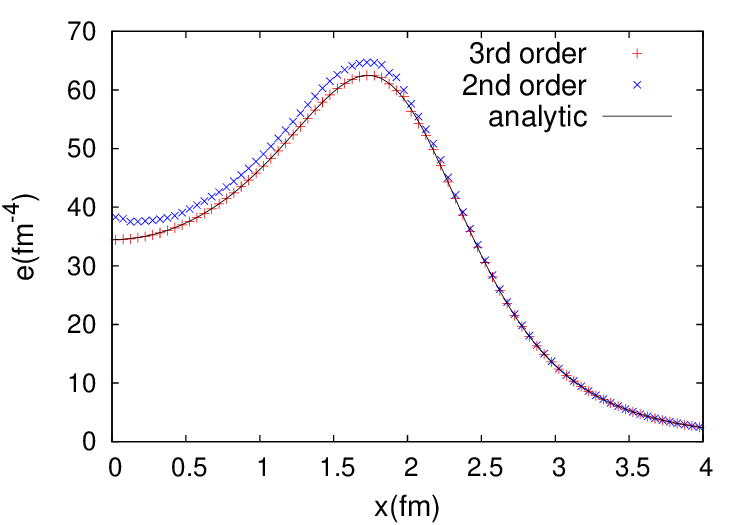}
  \centering
  \includegraphics[width=8cm]{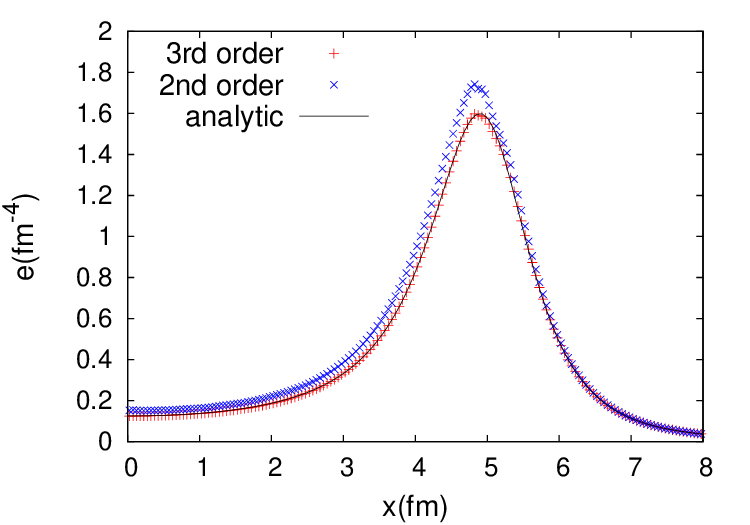}
  \caption{Comparison between 2nd order (MC limitter) and 3rd order (PPM) interpolation procedure 
  in the energy distributions. Top panel: At $\tau=2$ fm. Bottom panel: At $\tau=5$ fm. \label{fig:comparison_e}}
\end{figure}
\begin{figure}[b!]
  \centering
  \includegraphics[width=8cm]{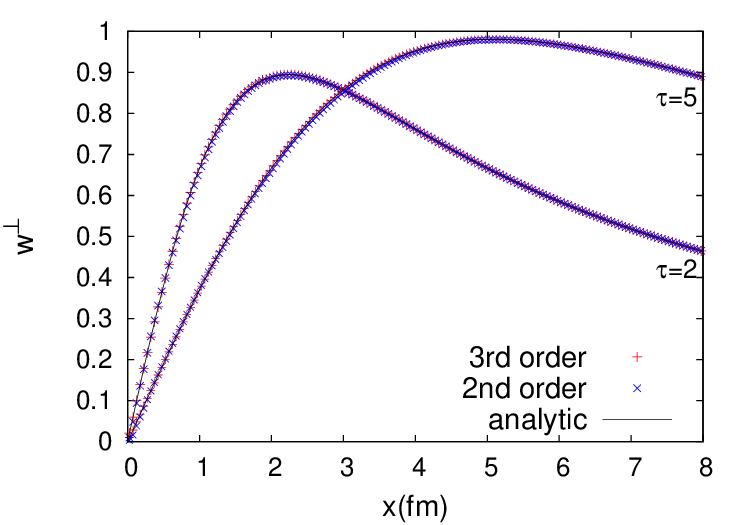}
  \caption{Comparison between the 2nd order (MC limiter) and the 3rd order (PPM) interpolation 
  procedure in $v_x$ at $\tau=2$ and 5 fm.
  \label{fig:comparison_vx} }
\end{figure}

\section{Importance of the interpolation scheme in Gubser flow}  \label{app-Gubser}

We discuss the importance of the interpolation procedure for the description of strong radial expansion 
like Gubser flow.  
Figure \ref{fig:comparison_e} shows the energy distributions at $\tau=2$ fm and 
$\tau=5$ fm which are obtained by the second- and third-order interpolation procedures. 
In the calculation, we use the same parameters as those in Sect.\:\ref{sec:Gubser}. 
The numerical calculations with the third-order interpolation procedure reproduce 
the analytical solutions. 
On the other hand, the energy density with the second-order interpolation procedure 
is slightly larger than that of the analytical solutions. 
In particular,  we observe the deviation from the analytical solution of energy density 
in $|x| < 2$ fm ($|x| < 5$ fm) at $\tau=2$ ($\tau=5$ fm). 
We find the same behavior in other second-order interpolation procedures, minmod 
and superbee limiters.
However, in the case of the one-dimensional expansion, even if the strong expansion 
exists, the numerical calculation with the second-order interpolation procedures shows 
good agreement with the analytical solution. 
We observe the deviation between the analytical solution and the numeral results 
in multidimensional calculation, 
which suggests that the operator splitting method is also a possible key issue for the problem. 

In Fig.\:\ref{fig:comparison_vx}, we show the transverse velocities at $\tau=2$ and $\tau=5$ fm. 
The gradient of the transverse velocity increases rapidly up to $x \sim 2$ ($x \sim 5$) fm at $\tau=2$ 
($\tau=5$) fm, where the value of the transverse velocity is slightly smaller than that of the analytical solutions, 
which implies that the second-order interpolation schemes do not satisfy the description 
of such a rapid expansion. The inadequate velocity growth causes the delay of the decrease of the energy 
density which is observed in Fig.\:\ref{fig:comparison_e}.

In Ref. \cite{Noronha2015} one points out the importance of adjusting the flux limiter 
in the algorithm (KT algorithm), using the relativistic viscous hydrodynamics. 
One shows that a free parameter $\xi$ in the van Leer minmod filter is fixed from a
comparison between the solutions of the shear-stress tensor from Gubser flow and the numerical 
calculations.




\end{document}